\documentclass{sig-alternate-2013}

\permission{Copyright is held by the International World Wide Web Conference Committee (IW3C2). IW3C2 reserves the right to provide a hyperlink to the author's site if the Material is used in electronic media.}
\conferenceinfo{WWW 2015,}{May 18--22, 2015, Florence, Italy.} 
\copyrightetc{ACM \the\acmcopyr}
\crdata{978-1-4503-3469-3/15/05. \\
http://dx.doi.org/10.1145/2736277.2741120}

\usepackage{color}
\usepackage[hyphens]{url}
\usepackage{verbatim} 
\usepackage{subfigure} 
\usepackage{multirow}
\usepackage{algorithm}
\usepackage[noend]{algorithmic}
\usepackage{xspace}
\usepackage{graphicx}
\usepackage{epsfig}
\usepackage{amsmath}
\usepackage{amssymb}
\usepackage{times}
\usepackage{float}
\usepackage{enumitem}
\usepackage{balance}
\usepackage[font=bf,belowskip=0pt,aboveskip=5pt]{caption}

\usepackage{booktabs} 

\newcommand{\sectionrule}{\addlinespace[0.5ex]}

\newcommand{\hide}[1]{}
\newcommand{\xhdr}[1]{\vspace{1.7mm}\noindent{{\bf #1.}}}

\newcommand{\DClong}{{DonorsChoose.org}\xspace}
\newcommand{\DC}{{DC.org}\xspace}
\newcommand{\etc}{\emph{etc.}}
\newcommand{\eg}{\emph{e.g.}}
\newcommand{\ie}{\emph{i.e.}}

\graphicspath{{./FIG/}}

\begin{document}

\title{Donor Retention in Online Crowdfunding Communities:\\A Case Study of DonorsChoose.org}

\numberofauthors{2}
\author{
\alignauthor
Tim Althoff\\
      \affaddr{Stanford University}\\
      \email{althoff@cs.stanford.edu}
\alignauthor Jure Leskovec\\
      \affaddr{Stanford University}\\
      \email{jure@cs.stanford.edu}
}

\maketitle

\begin{abstract}

Online crowdfunding platforms like DonorsChoose.org and Kickstarter allow specific projects to get funded by targeted contributions from a large number of people. 
Critical for the success of crowdfunding communities is recruitment and continued engagement of donors. 
With donor attrition rates above 70\%, 
a significant challenge for online crowdfunding platforms as well as traditional offline non-profit organizations is the problem of donor retention.

We present a large-scale study of millions of donors and donations on DonorsChoose.org, 
a crowdfunding platform for education projects. 
Studying an online crowdfunding platform allows for an unprecedented detailed view of how people direct their donations. 
We explore various factors impacting donor retention which allows us to 
identify different groups of donors and quantify their propensity to return for subsequent donations.
We find that donors are more likely to return if they had a positive interaction with the receiver of the donation.
We also show that this includes appropriate and timely recognition of their support as well as detailed communication of their impact.
Finally, we discuss how our findings could inform steps to improve donor retention in crowdfunding communities and non-profit organizations.

\end{abstract}

\noindent {\bf Categories and Subject Descriptors:} H.2.8 {\bf
[Database Management]}: Database applications---{\it Data mining}


\noindent {\bf Keywords:} Donor Retention; User Retention; Crowdfunding. 

\section{Introduction}
\label{sec:intro}

Crowd-sourced fundraising, or {\em crowdfunding}, for short, provides a revolutionary way for organizations and projects to collect funding. 
Online crowdfunding platforms such as Kickstarter.com or \DClong allow individuals to post project requests in order to raise funds for the development of new products, to support artistic and scientific endeavors, and to contribute to public education~\cite{Mollick14Dynamics,wash2013value}. 
Anyone can become a donor and direct small contributions to specific projects and this way, the ``crowd'' collectively contributes to the funding of the project.
Even though, projects solely rely on contributions from a large number of individuals, 
crowdfunding projects have raised over \$2.7 billion in 2012 alone~\cite{MacLellan2012CrowdfundingRise}.

A critical component for the success of fundraising campaigns is the recruitment of new and engagement of existing donors. {\em Donor retention} refers to the problem of keeping donors that continue to make donations year after year.

Present donor retention rates are only around 25\% for first time donors~\cite{fep2013fundraisingreport,sargeant2008donor} and increasing donor retention would have significant impact on the effectiveness of online as well as offline fundraising campaigns.  First, it can be much more cost-effective to maintain relationships with existing donors than to recruit new donors. And second, even small improvements in donor retention can have a significant impact on the amount of collected funds. 
For example, a 10\% improvement in donor retention could yield up to a 200\% increase in obtained donations~\cite{sargeant2008donor}.

Despite the importance of donor retention for fundraising campaigns, many of its basic aspects are still not well understood. Current knowledge about donor retention largely consists of anecdotal evidence from fundraising professionals and small lab experiments in artificial environments (for a survey, see~\cite{sargeant2008donor}). 
There are many questions about donor retention that remain open. For instance, 
are different donor subgroups affected differently by timely acknowledgments?
What does timely even mean and what can we infer about the donor's expectations from their behavior?


\xhdr{Present work: Donor retention in online crowdfunding communities}
In this paper, we study the {\em intersection} of crowdfunding communities and charitable organizations by studying an online charity that allows donors to donate to very specific small projects of their choosing (\ie, operating exactly like a crowdfunding platform): \DClong (\DC). 

We focus on the problem of donor retention as it is a fundamental problem both for online crowdfunding platforms as well as to a large and rapidly growing sector of non-profit organizations and charities~\cite{fep2013fundraisingreport,barber2013donor,sargeant2008donor}.



\begin{figure}[t]
  \centering
  \includegraphics[width=1.0\linewidth]{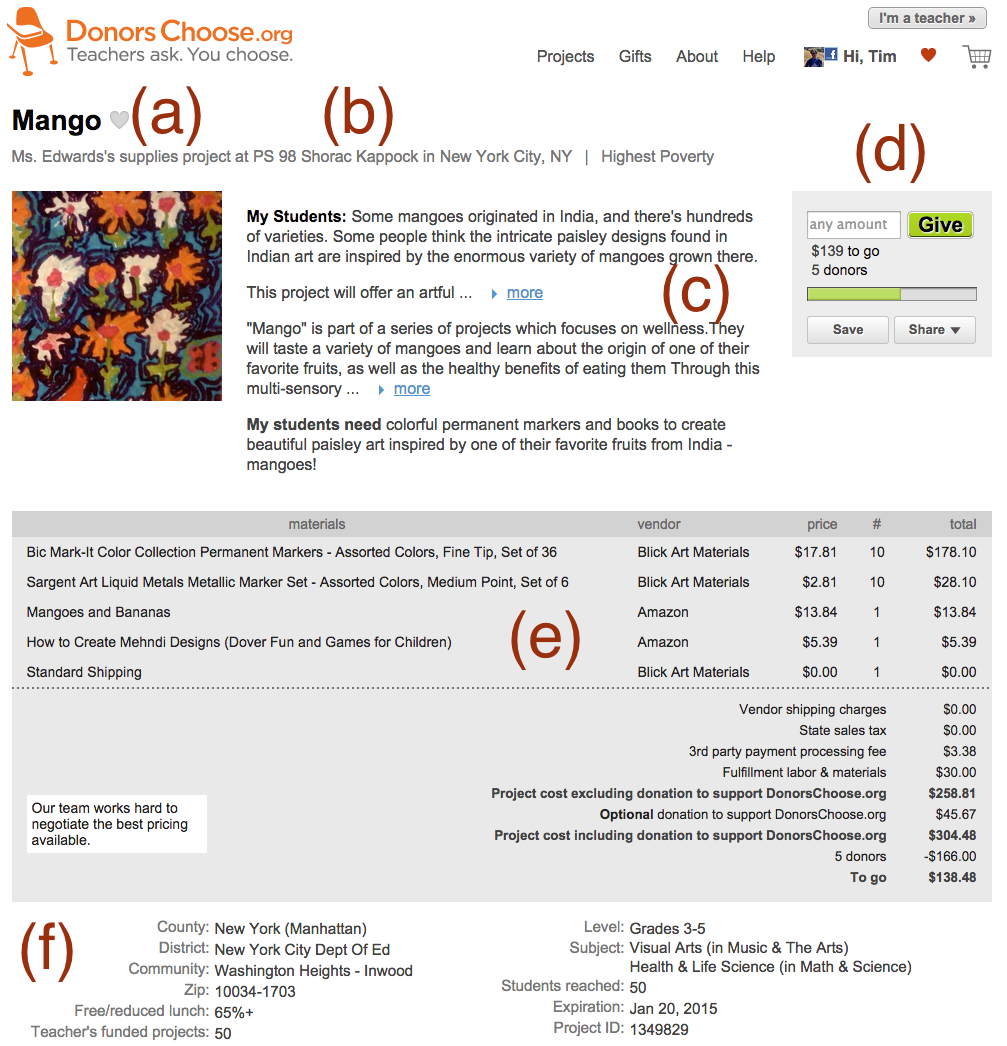}
  \caption{
    Example project request from \DClong.
    Every project page contains the project title (a); 
    the teacher and school (b); 
    an essay by the teacher about their students, their project, and their specific need (c);
    the remaining amount to fund the project and the number of donors who have given already (d);
    needed materials in itemized form (e);
    and more information about the school and its students (f).
    This project asks for art supplies for a primary school in New York City.
    }
  \label{fig:dc_project_screenshot}
\end{figure}

We analyze a complete trace of donor and project activity from \DClong, a U.S. nonprofit organization that allows teachers to easily post requests for donations to purchase materials in  support of their classroom. Through \DC, teachers compose a short essay on their students and project plans and itemize needed materials. An example project is shown in Figure~\ref{fig:dc_project_screenshot} in which an elementary school teacher in a high-poverty district of New York City asks for ``\$305 to purchase colorful permanent markers and books to create beautiful paisley art inspired by one of their favorite fruits from India --- mangoes!''

Our data contains complete project activity from the inception of \DC in March 2000 to October 2014. In this time, \DC attracted over 1.5M donors, 638k projects, and 3.9M donations for a total of \$282M. More than 60\% of all public schools in the U.S. have raised money for their classrooms through \DC to date~\cite{donorschoose2014impactpage}.

To the best of our knowledge the present work is the first study of donor retention in online crowdfunding platforms.

\xhdr{Summary of results}
Online crowdfunding platforms face the same problem of donor retention as traditional non-profit organizations.
We show that only 26\% of first-time donors ever return and donate a second time.
Thus, increasing donor retention on \DC would have huge impact that tens of thousands of public schools could benefit from.
We start addressing this issue by analyzing the donation behavior of first-time donors, for which the attrition rate is largest.
We identify a set of factors related to donor retention in the context of \DC. 
We study these factors empirically and quantify their effect on donor retention.
We find that factors such as entering the community through different means, geographical patterns of donations, the donation amount, and disclosure of optional personal information all shine light on the donor's initial motivations and signal commitment to the crowdfunding community. 
Furthermore, factors including project cost, project success, and timely responses by the teacher (highlighting the impact the donor has made) can affect the donor's sense of personal impact and trust in the organization (which are known to positively influence retention in offline charities~\cite{sargeant2008donor}).
In addition, we show that the teacher's ability to retain donors is correlated with their experience, timely writing thank you notes, and the use of Facebook for solicitation.

We also show that whether a donor will return for a second donation can be predicted just based on the properties of the donor's first donation. We build a machine learning model to predict donor return on an individual level with promising accuracy.
Finally, we discuss how these results could be translated into actionable suggestions for online crowdfunding communities as well as traditional offline non-profits and charitable organizations.


\xhdr{Outline}
The rest of the paper is structured as follows.
Section~\ref{sec:dataset} introduces DonorChoose.org, the dataset, and donor retention within the community.
The analysis of retention factors is split into three perspectives: 
project (Section~\ref{sec:project}),
donor (Section~\ref{sec:donor}), and
teacher (Section~\ref{sec:teacher}).
We provide a brief summary (Section~\ref{sec:summary}) before demonstrating that donor retention can be predicted on an individual level (Section~\ref{sec:prediction}).
We describe related work in Section~\ref{sec:related}
and conclude with a discussion and future work in Section~\ref{sec:conclusion}.


\section{Dataset Description}
\label{sec:dataset}

This section gives details on the mechanics of the \DC crowdfunding platform, 
the dataset used in this paper,
and a first look at the state of donor retention on \DC.

\subsection{The Mechanics of \DClong}
\label{sec:mechanics}
\DC enables {\em teachers} to request materials and resources for their classrooms and makes these project requests available to individual {\em donors} through its website  (teachers can act as donors, too).
In contrast to most offline non-profit organizations (NPOs), {\em projects} on \DC are very concrete and provide an itemized list of the materials they ask for.
In this regard, \DC is more similar to other crowdfunding platforms.
{\em Project pages} contain an essay by the teacher and further information about the concrete need, the school, location, poverty level, subject, grade level, how many students are reached by this project, and how many projects by the teacher have been successfully funded in the past (see Figure~\ref{fig:dc_project_screenshot}).
If a partially funded project expires (\ie, fails to attract full funding within a four month period), donors get their donations returned as account credits, which they can use towards other projects.
When a project does get fully funded, \DC purchases the materials and ships them to the school directly.
At this point the teacher will send a so-called {\em confirmation note} to all donors thanking them for their donations.
After the materials arrive in the teacher's classroom, the teacher will compose an {\em impact letter} giving insights on how the donor's support has made an impact in their classrooms. 
Often, these impact letters will come with photos of students using the donated materials.
Donors who contribute \$50 or more to a project can also request hand-written {\em thank you notes} from the students.

Donors can enter the site through different means.
Some enter through the \DC front page and use the search interface to find projects they feel passionate about (allowing them to filter or sort by many attributes including school name, teacher name, location, school subject, school material requested, keywords, cost, \etc).
By default, the interface sorts projects by urgency (high poverty and close to finish line) and displays those projects at the top of the page.
There is no further personalization, i.e. all visitors to the site see the same projects.
We will refer to these donors as {\em site donors} for the rest of this paper.
Other donors are referred by the teacher to make a donation to their own project on \DC.
For example, teachers often share the project URL with friends, family, parents of their students, or through posts on social media.
\DC tracks how donors enter the site and attributes any resulting donations to the teacher's fundraising efforts.
We will refer to this group of donors as {\em teacher-referred}.
Donors give about \$55 dollars on average per donation (minimum \$1, median \$25).

\subsection{The Dataset}
We use a complete dataset of all projects and donations to \DC from their inception in March 2000 to October 2014.
Our dataset contains 3.9M donations by 1.5M donors to 638k projects over a total amount of \$282M.
We restrict our analysis to donations after Jan 1, 2009. 
Before then \DC was relatively small and only started operating nationwide in 2008; also, the website interface has changed over time.
In all our analyses we filter out grant accounts held by special partners, promotions, gift card purchases that do not go towards a specific project, all donations by teachers, and all donations that are fully paid by account credit (\ie, we require the donor to spend actual money).


\subsection{Donor Retention on DonorsChoose.org}
\label{subsec:donor_retention_on_dc}
\begin{figure}[t]
  \centering
  \includegraphics[width=1.0\linewidth]{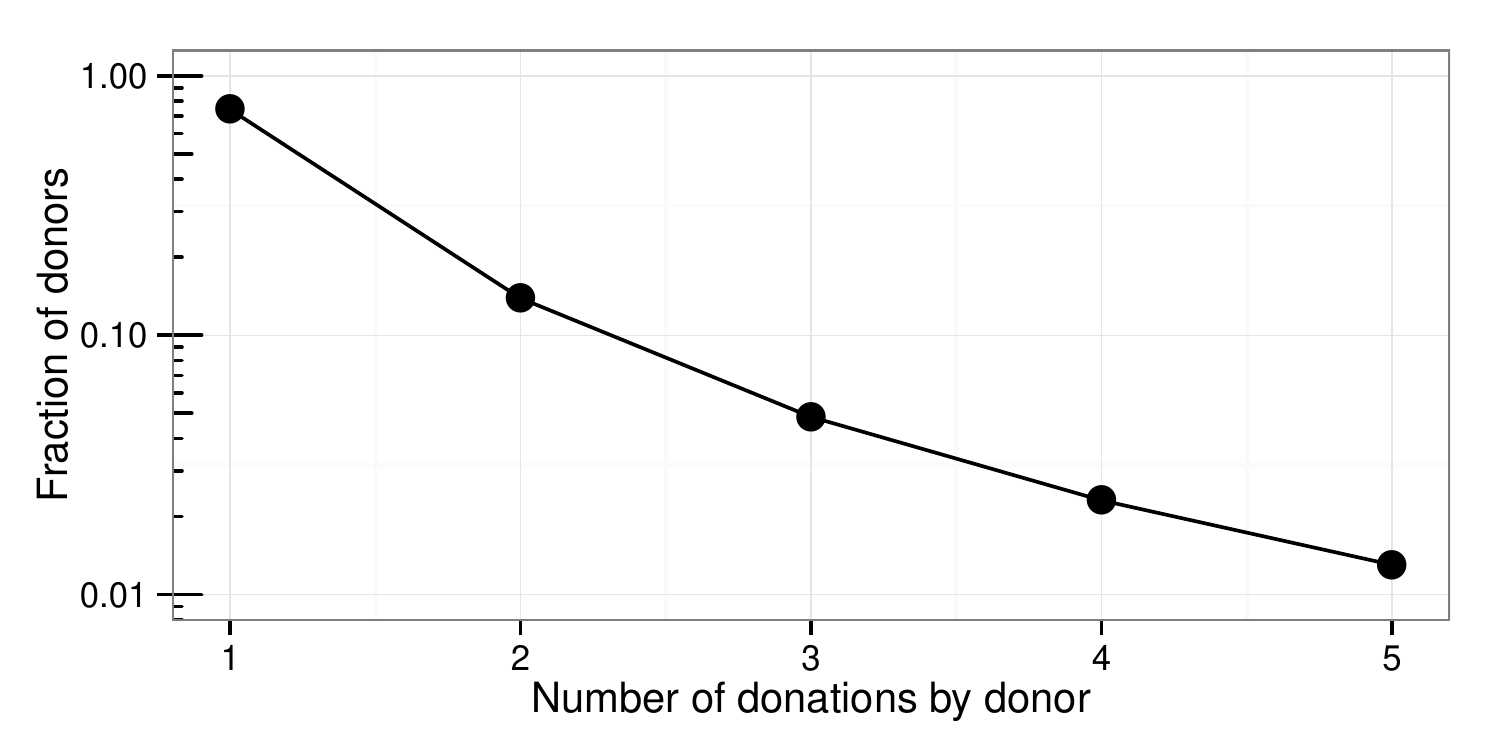}
  \caption{
    Fraction of donors by number of total donations by the donor.
    Note the log scale of the Y axis.
    74\% of donors donate exactly once and do not return. 
    26\% return for a second donation and 14\% do not return afterwards. 
    Only 1\% of donors make five donations.
    Online crowdfunding is facing the same attrition problem as offline NPOs.
        }
  \label{fig:donor_attrition}
  \vspace{5mm}
\end{figure}

In this paper, we focus on donors and ask the question which donors return to make another donation.
We identify that of all donors that do return over the course of our observation period, most do so within one year. 
Therefore we give every donor at least one year to return (\ie, we only look at donations until September 2013 and use the following year to determine whether the donor returned or not).
The presented results are robust to slight variations of this definition.

Figure~\ref{fig:donor_attrition} shows what fraction of donors makes how many donations within the entire observation period: 74\% of donors make exactly one donation and never return, 14\% return for a second donation, and only 1\% of all donors make five donations (over a five year period).
Attrition is highest and most potential is lost after the first donation and this is where we focus our attention in this paper
(the importance of obtaining a second donation for donor retention has also been recognized recently in \cite{love2014firsttimedonors}).

Thus we focus our analyses on a set of 470k first-time donors of which 26\% returned for a second donation.
We also analyze return to the same teacher (as opposed to the overall site) for which we additionally require that the second donation went to the same teacher as the first donation (more in Section~\ref{sec:teacher}).
Our dataset is representative of the issues the field is facing with very low retention rates.
We observe a negative trend over time similar to what has been reported for offline NPOs \cite{barber2013donor}. In particular, in \DC the donor retention for first-time donors fell from about 35\% in 2009 to under 25\% in 2013.

We structure our following analysis of donor retention factors into three parts 
focusing on projects (Section~\ref{sec:project}), donors (Section~\ref{sec:donor}), and teachers (Section~\ref{sec:teacher}).

\section{Project Perspective}
\label{sec:project}
Many factors around the donor's first interaction with \DC are related to donor retention.
This section focuses on the subset of factors around the project that the donor supported with their first donation.
In particular, we study the effect of project success and project cost on donor retention.

\pagebreak 
\subsection{Trusting the System -- Project Success}
\label{subsec:project_success}

\begin{figure}[t]
  \centering
  \includegraphics[width=1.0\linewidth]{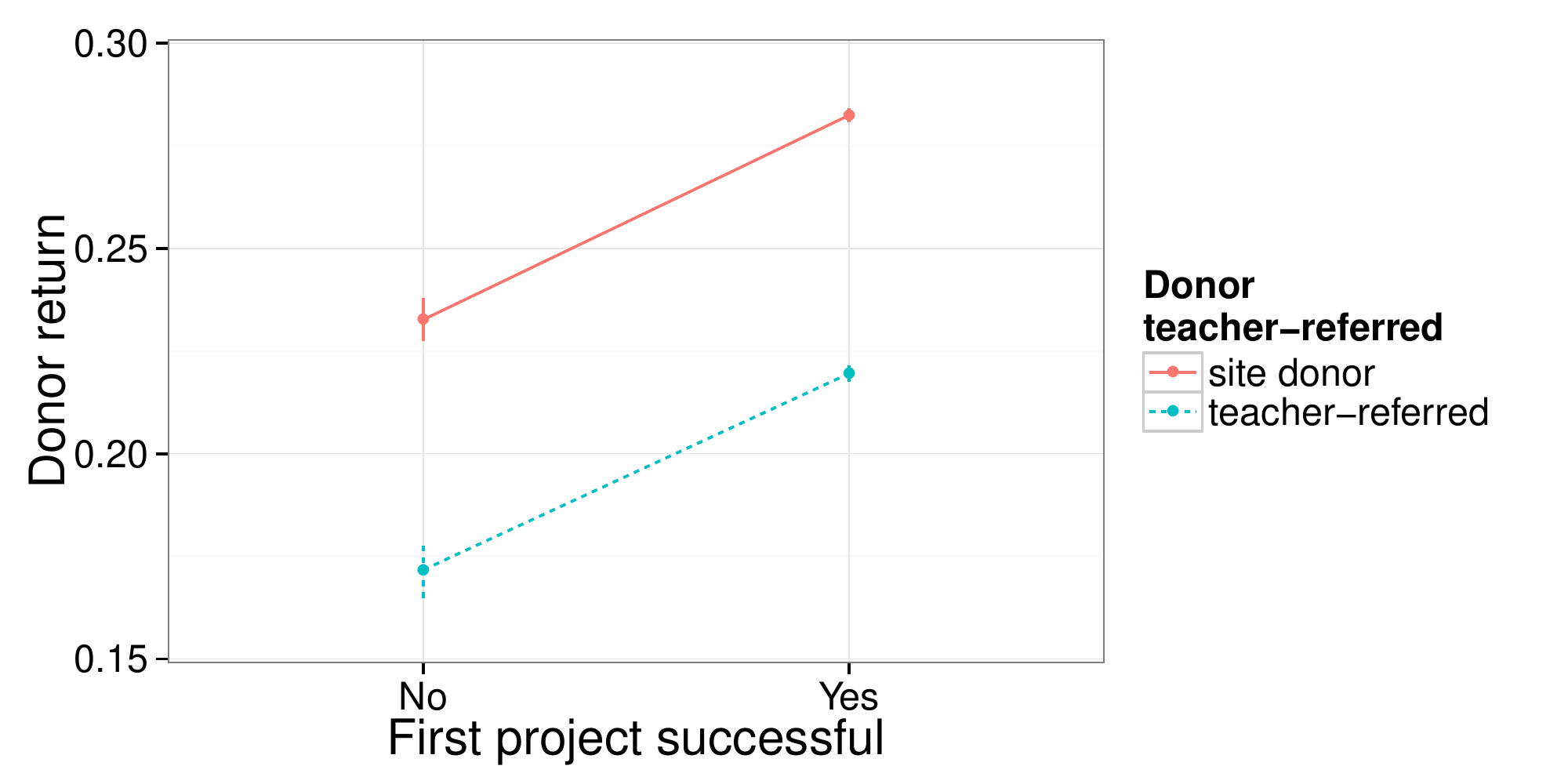}
  \caption{
    Whether or not the first project is successful is strongly correlated with retention
    for both teacher-referred and site donors.
    Donors whose first project succeeded are 5\% more likely to return and donate again.\hspace{\columnwidth}
    Note: The error bars in all plots represent 95\% confidence intervals on the corresponding mean estimate.
    }
  \label{fig:return_rate_by_project_success}
\end{figure}

Trust between donor and organization has been identified as a key driver of loyalty \cite{sargeant2008donor}. 
%
Arguably, project failure indicates inability (of the site or teacher) to use the donated funds successfully towards the shared goal of improving public education.
Therefore, we might expect to see higher return rates among donors whose first project was successfully funded compared to return rates of donors that were unsuccessful initially.
The results are shown in Figure~\ref{fig:return_rate_by_project_success}.
We find that first-time donors are about 5\% more likely to return if their first project is successfully funded.
This means that when a person donates to a project that fails to attract enough funds, that person is less likely to make another donation.
The effect is substantial if one considers the low baseline rates (a relative increase of 29\%) and the fact that even small retention increases can have a high impact on total donations~\cite{sargeant2008donor}.

However, the finding could be confounded by the fact that successful projects are likely to ask for a smaller total amount and receive higher than average donations (and we find they do). 
To control for the effect of these confounders we use an almost-exact matching strategy \cite{rosenbaum2010observational} 
in which we pair donations that are identical in donation size (less than one cent difference on average), project cost (less than one dollar difference on average), and a variety of other factors (donation included optional support, grade level, poverty level of school district, being teacher-referred). 
Performing the analysis on 20k matched pairs of donations we find the same effect, albeit slighty reduced (3.2\% increase).

To sum up, we conclude that donors who donate to a successful project are substantially more likely to return. The observation could be explained by several factors: donors trust the site more as their donation actually gets used, and also by donating to the successful project donors might get a greater sense of impact. It is exactly the impact that we examine next.

\subsection{A Personal Sense of Impact -- Project Cost}
\label{subsec:project_cost}
Most projects on \DC ask for amounts between \$200 and \$600.
Even though \DC instructs teachers that smaller projects are more likely to be successful, some teachers ask for more than \$1500 (\eg, for expensive computer equipment).
Such projects usually require more donors (often several dozen) to get funded. 
In this context, an individual donor might have less of a sense of personal impact.
If they are, say, one of a small number of donors that fully funded the project they might have a greater sense of accomplishment and impact than if they are one of many donors to a project~\cite{sargeant2008donor}.

\begin{figure}[t]
  \centering
  \includegraphics[width=1.0\linewidth]{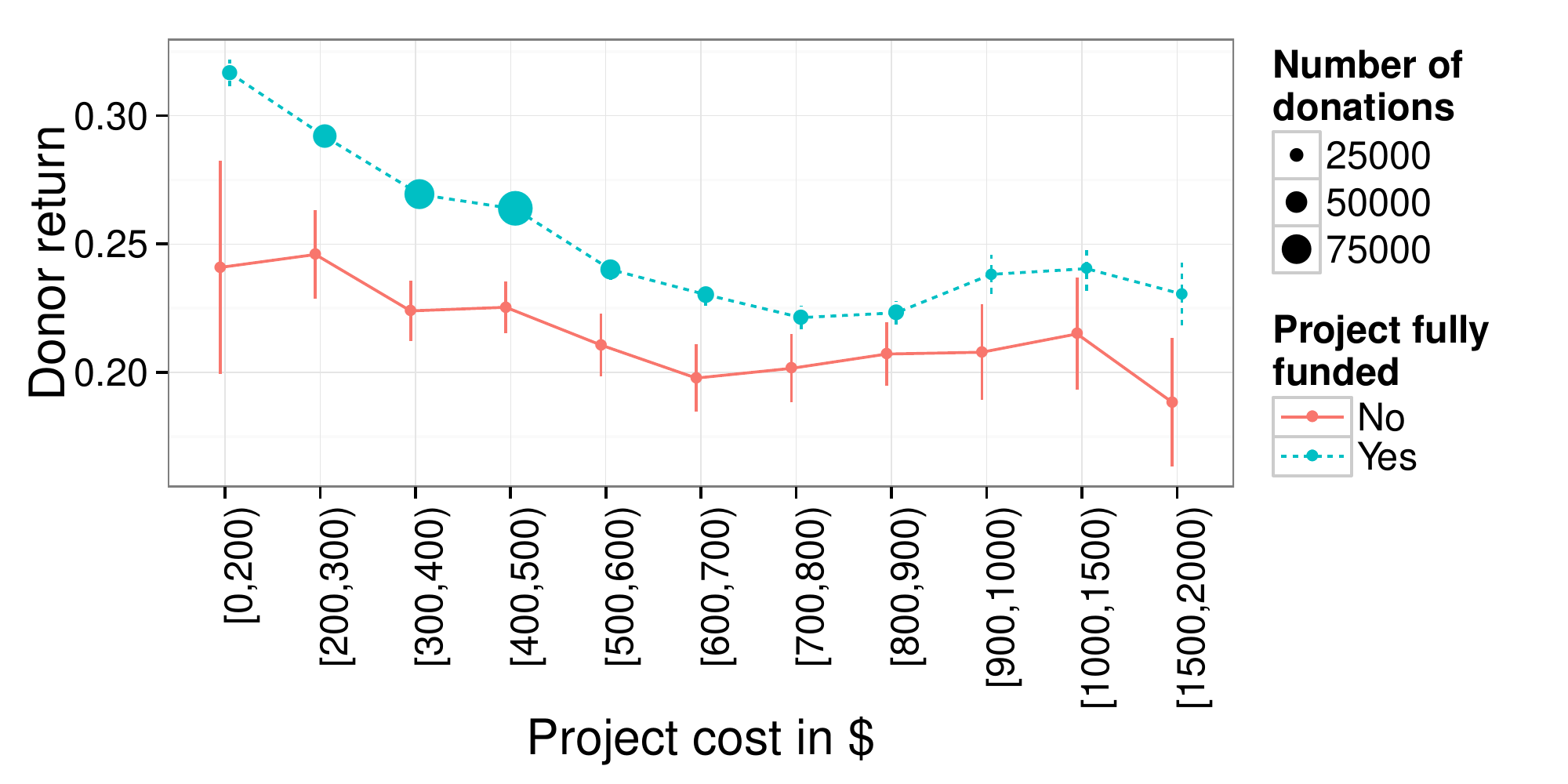}
  \caption{
    First-time donors to small projects are much more likely to return than donors to large projects.
    This effect could be explained through having a greater sense of personal impact when fewer people make the project succeed.
    }
  \label{fig:return_rate_by_project_cost}
\end{figure}

We show a graph of return rates as a function of project cost in Figure~\ref{fig:return_rate_by_project_cost}. The graph shows two curves: donor return for fully funded projects (blue) and donor return for non-funded projects (red). 
To give a sense for the donation distribution, the area of the circles is proportional to the number of donations to a project of a given amount.

We make several observations.
First, the blue curve (successful projects) is always above the red curve (unsuccessful projects), which is consistent with the previous finding (Section~\ref{subsec:project_success}) that donor return rate is positively correlated with the project success across all project sizes.
Second, we observe that donors to small successful projects are much more likely to return (32\%) than donors to large projects (23\%).\footnote{
Note that most differences in mean return rate for successful projects are statistically significant as the 95\% confidence intervals (error bars) are disjoint.}
For projects larger than \$600 we observe no difference.
And third, we find find the same effect for donors that make small donations as well as donors that make large donations (not shown in figure; more about donation sizes in Section~\ref{subsec:donation_amount}).


\section{Donor Perspective}
\label{sec:donor}
Next, we explore retention factors around the donors themselves
and the first donation they make.

\subsection{Tracking How Donors Joined \DC\\-- Teacher-referred Donors}
\label{subsec:teacher_referred_donors}
Considering how a particular donor found out about \DC can shine light on their personal motivations and even whether they might know the teacher who started the project personally.


We split first-time donors into two groups {\em teacher-referred} donors and {\em site} donors (see Section~\ref{sec:mechanics}).
These two groups of donors are arguably quite different. 
While donors of the first group presumably entered the community to support a specific teacher, the donors of the second one could have joined the community because they wanted to support the cause of improving public education and/or the community as a whole.

\begin{figure}[t]
  \centering
  \includegraphics[width=1.0\linewidth]{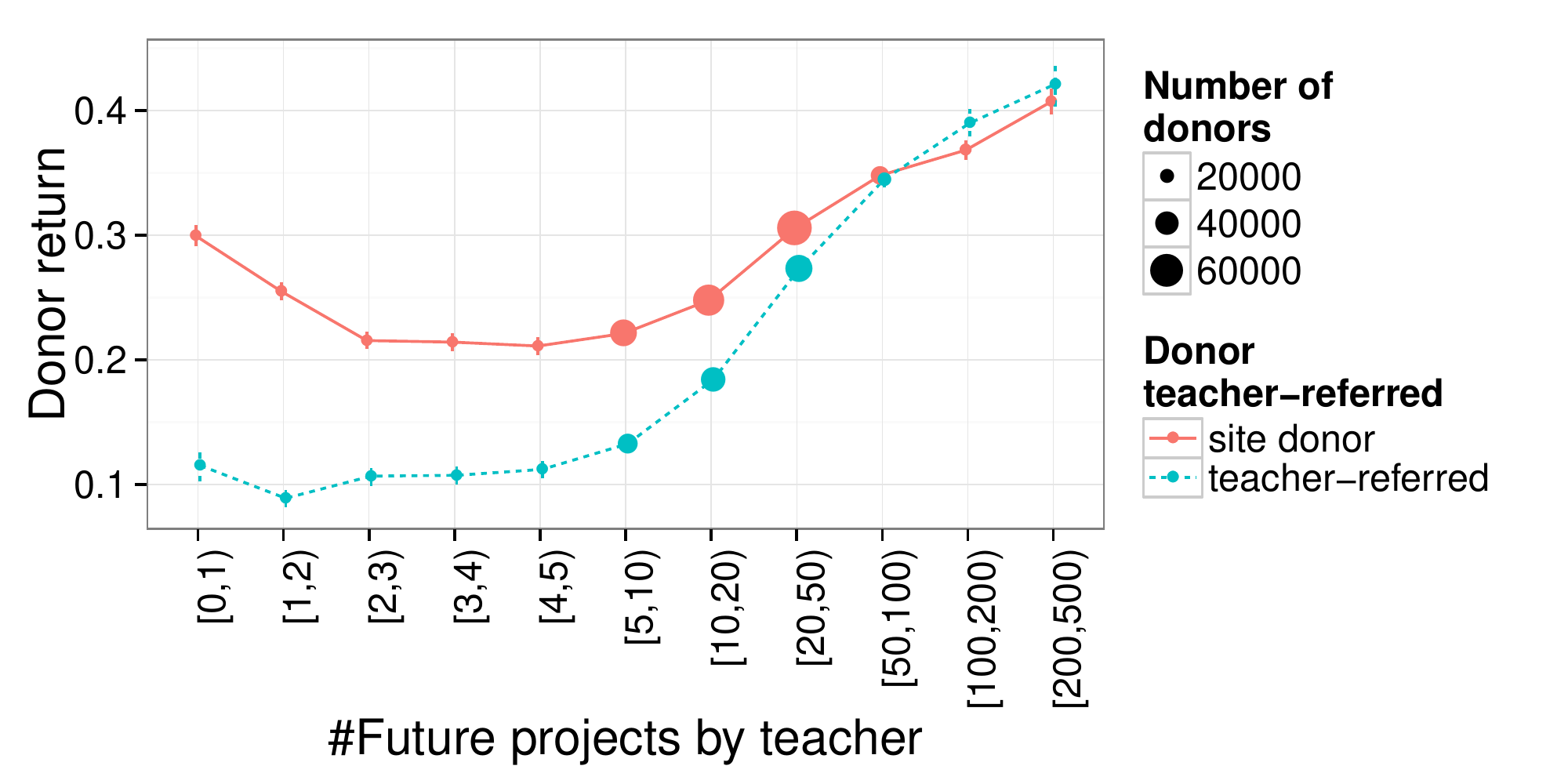}
  \caption{
    The propensity to return as a function of the number of projects by the same teacher in the future.
    Teacher-referred donors (blue) are less likely to return than site donors (red).
    Both types of donors are much more likely to return if they donate to (future) ``high-profile'' teachers.
    }
  \label{fig:return_rate_by_n_future_projects_teacher}
\end{figure}

We show retention rates for both groups in Figure~\ref{fig:return_rate_by_n_future_projects_teacher}.
The plot shows the propensity to return as a function of the number of future projects by the same teacher (of the first project donated to).
The distinction is relevant as it controls for the amount of future solicitation by the teacher who is likely to reach out to previous donors again when he or she starts a new project.
Overall teacher-referred donors are less likely to return than site donors (blue vs. red curve).

We notice that for teachers that only create very few additional projects, the return rates are very different between the two groups.
The non-teacher-referred donors (site donors; red) arguably have higher intrinsic motivation to continue donating whereas teacher-referred donors (blue) lost their main reason to be part of the crowdfunding community and thus rarely return.
However, the more projects the teacher will start in the future (\ie, the more often the teacher returns) the more likely are teacher-referred donors to return.
Likely this is due to increased solicitation efforts by that teacher.
Interestingly, the return rate also increases for non-teacher referred donors. 
This could be explained by the fact that teachers are likely to reach out to these donors for future projects.
In addition, \DC might notify both kinds of users of future projects as part of their recommender system.


\subsection{Distance as a Proxy for Involvement}
\label{subsec:distance}

\begin{figure}[t]
  \centering
  \includegraphics[width=1.0\linewidth]{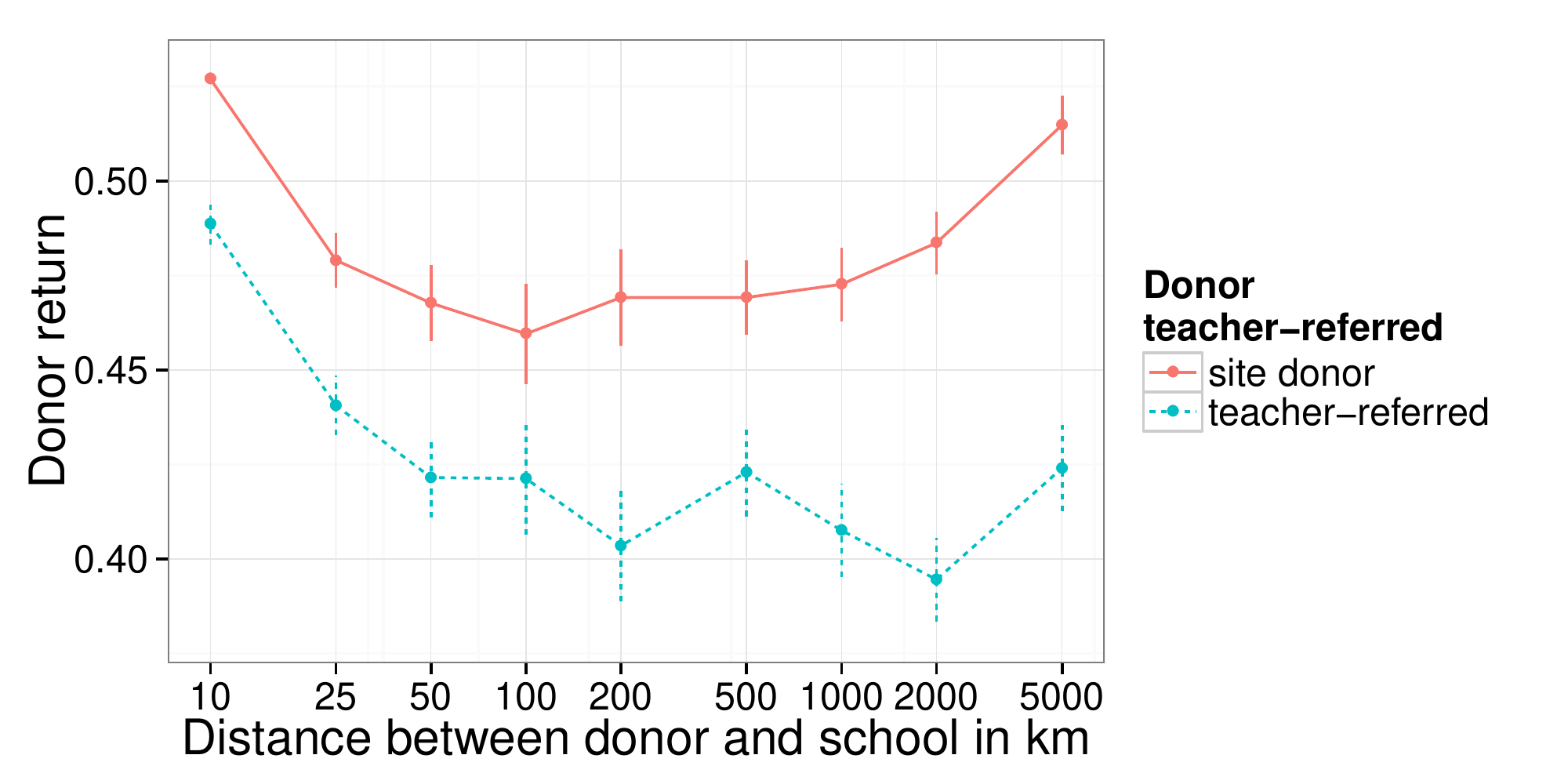}
  \caption{
    Retention rate across donors located at varying distances from the projects they funded with their first donation.
    Local donors are generally more likely to return and distant {\em site} donors are particularly loyal to \DC as well.
    }
  \label{fig:return_rate_by_distance_and_teacher_ref}
\end{figure}

On \DC donors can specify their location and zip code in their user profile. In addition, all projects are highly geo-specific as they are funding a particular classroom within a particular school at a particular location.
%
%
Having location information on both donors and projects allows us to put these in relation and measure how far donors live from the project and then explore how this is correlated with donor retention. 


We form groups of donors based on the distance between them and the school they funded with their first donation (using the center point of their zip code and latitude/longitude coordinates for the school).
The return rates within those groups are shown in Figure~\ref{fig:return_rate_by_distance_and_teacher_ref}, separately for teacher-referred and site donors. 
The $x$-axis labels correspond to the upper bound on the distance interval of that group, for example ``25'' corresponds to all donors of the (10km, 25km] range.
Again, we observe that teacher-referred donors are less likely to return across all distance groups.
We find that local donors within 25km are most likely to return for both groups.
As discussed before, these are the donors that could be more likely to be personally involved with the school or particularly care about impacting their local community.
Interestingly, the retention rate increases again for distant site donors, forming a ``U-shape''. 
Site donors that live across the country (\DC is a solely USA-based community) are almost as likely to return as their local counterparts.
This subpopulation represents donors that are very passionate about the community and overall cause that they will even fund projects across the country when they (most likely) do not know the teacher.
This effect is not as present for teacher-referred donors where the retention rate seems to plateau for distant donors.

Note that the above analysis is constrained to donors that share their location.
Also note that the average return rate in Figure~\ref{fig:return_rate_by_distance_and_teacher_ref} is significantly higher than the overall average (26\%).
Whether or not a donor gives away their location is a signal in itself that we analyze in Section~\ref{subsec:disclosing_personal_information}.
We will also see in Section~\ref{subsec:donation_position} that local giving is a large driver of donation volume.

\subsection{The Donor's Role within the Project\\-- Donation Position}
\label{subsec:donation_position}
Next, we investigate what we can learn about the donor based on the role they are assuming in the project they fund through their first donation.
This is connected to the concepts of self-definition and identificiation of the donor that the fundraising literature has identified als influencers of donor loyalty~\cite{sargeant2008donor}.

We hypothesize that donors might fall into three categories: 
{\em starters} that like to start off new projects with an initial donation,
{\em closers} that like to finish off projects that are close to completion,
and a third group that does not particularly follow any of the previous two behaviors.
How likely are these three groups to return?

\begin{figure}[t]
  \centering
  \includegraphics[width=1.0\linewidth]{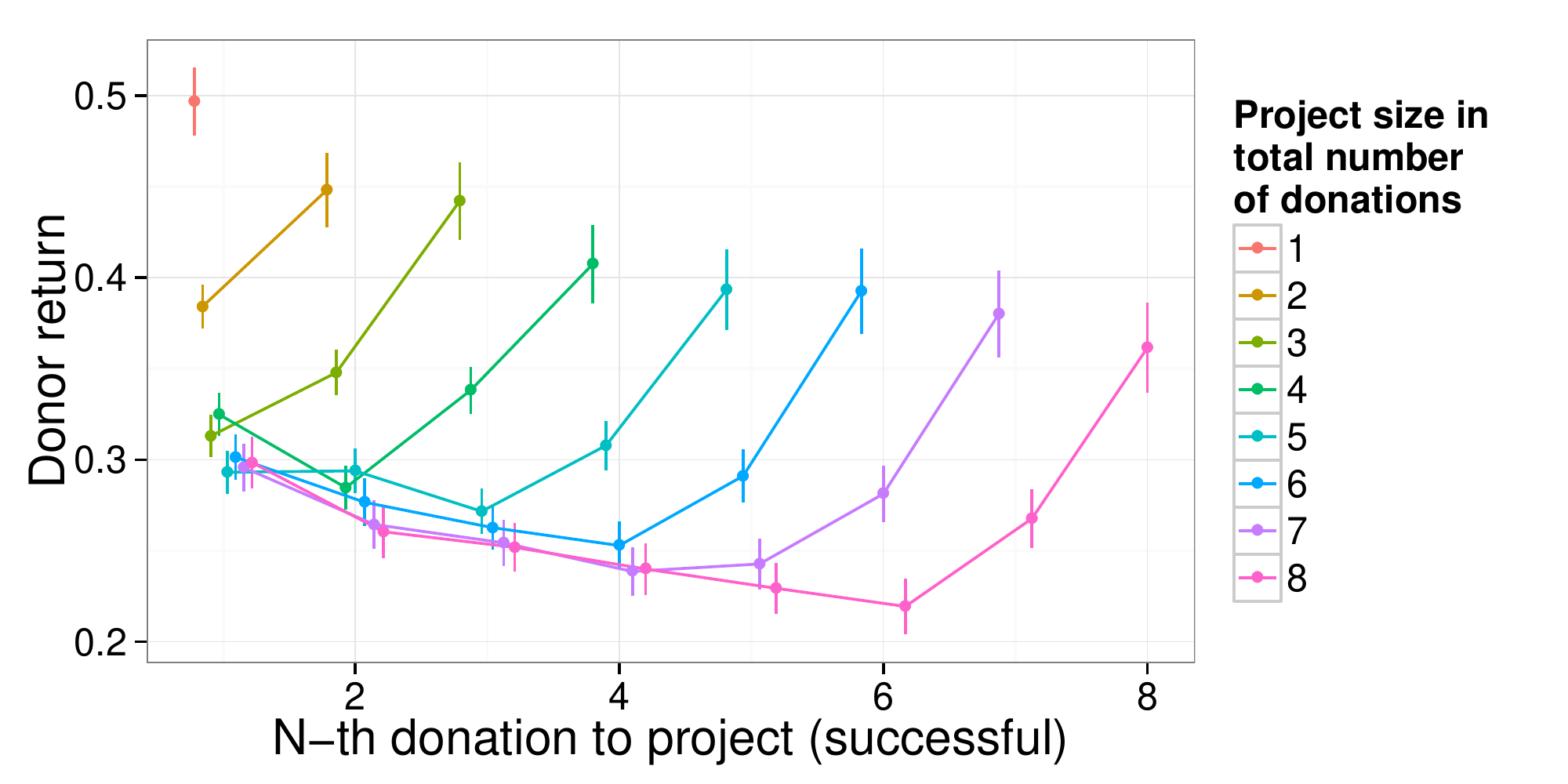}\\
  \includegraphics[width=1.0\linewidth]{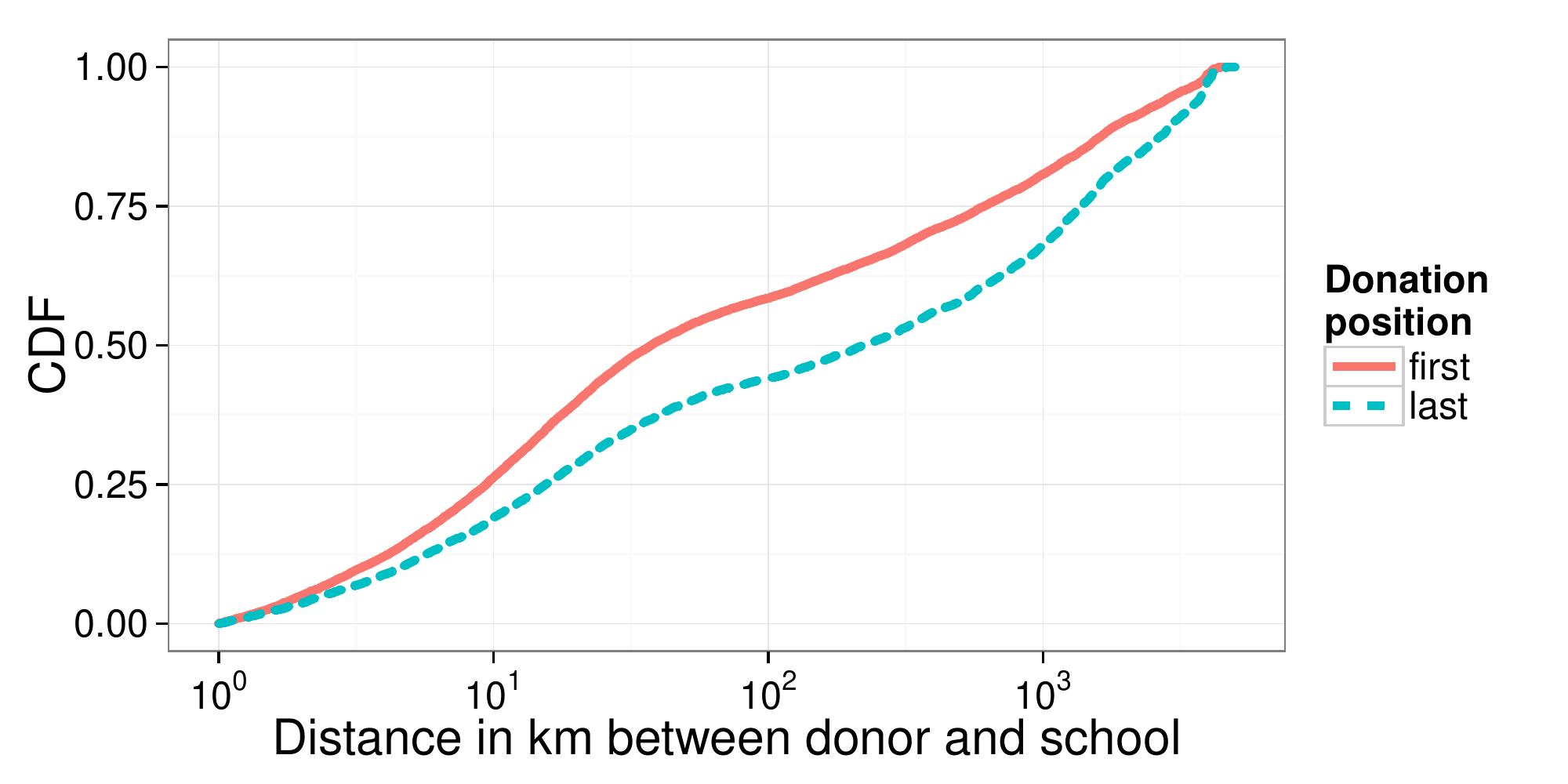}
  \caption{
    Top:
    Donors that donate last to projects are much more likely to return than other donors (for successful projects).
    Early donors are more likely to return than middle ones.    
    Across all project sizes we observe a consistent ``U-shape''.
    Bottom:
    Early donors tend to be local and late donors tend to be distant.
    }
  \label{fig:return_rate_by_nth_don_to_project}
\end{figure}

Figure~\ref{fig:return_rate_by_nth_don_to_project} (top) shows donor retention across donation position for successful projects that received between one and eight total donations.
We observe a remarkably consistent ``U-shape'' trend across all project sizes in which donors in the middle of the project's lifetime are less likely to return than the starters. Moreover, closers, by far, display the highest propensity to return for another donation.

It seems unlikely that donation position by itself would have such a strong effect on the return rate (although it is conceivable that starters or closers feel a greater sense of impact when the project succeeds). More likely, these are actually different groups of users that interact with the project at different points of its lifetime.

Let us attempt to understand how these groups of donors might be different.
Consider Figure~\ref{fig:return_rate_by_nth_don_to_project} (bottom) that shows the distance distribution (cumulative distribution function; CDF) for first-time donors that make the first donation to their project (starters in red) and and for first-time donors that make the last donation to their project (closers in blue).
We find that starters are more likely to be local and that closers are more likely to be distant (consistent with findings in~\cite{agrawal2011geography}).
From Section~\ref{subsec:distance}, we know that local as well as very distant donors are more likely to return, which leads to the U-shape in Figure~\ref{fig:return_rate_by_distance_and_teacher_ref}.
Early donors are also more likely to be teacher-referred,
this group is also generally less likely to return (Figure omitted due to space constraints). 
This partially explains the large effect and U-shape in Figure~\ref{fig:return_rate_by_nth_don_to_project} (top).
Figure~\ref{fig:return_rate_by_nth_don_to_project} (bottom) also shows that local giving is very important on \DC as donations from within a 10km neighborhood around the school make up to 28\% of first donations to projects and donations from within 100km make up about 50\% of all donations
(of donors that specified their location).
Note that this effect is not due to the user interface or personalization as projects are not sorted by distance (which could favor local projects; see Section~\ref{sec:mechanics}).


\subsection{Are You Committed to \DClong?\\-- Donation Amount}
\label{subsec:donation_amount}
This section analyzes the relationship between donor return and the donation amount of the donor's first donation.
Arguably, giving large amounts of money is one way to display high levels of commitment (though we recognize that less affluent donors can be committed as well).
In the absence of a direct measure of commitment, we are therefore using donation amounts as a proxy.
The fundraising literature defines commitment to an organization as the donor's desire to maintain a relationship or,
alternatively, a genuine passion for the future of the organization and the work it is trying to achieve.
Since commitment is positively correlated with loyalty~\cite{sargeant2008donor}
we would expect that first-time donors giving larger amounts are more likely to return in the future.


\DC raises money to support their organization by asking for an optional support with each donation.
By default, 15\% of each donation goes towards \DC and most people (85\%) include this optional support for the site/organization.
This gives us another signal of how committed donors are to the organization---if they explicitly opt out of supporting \DC we would expect that they care more about the particular project or teacher they are supporting than they care about \DC more generally.

\begin{figure}[t]
  \centering
  \includegraphics[width=1.0\linewidth]{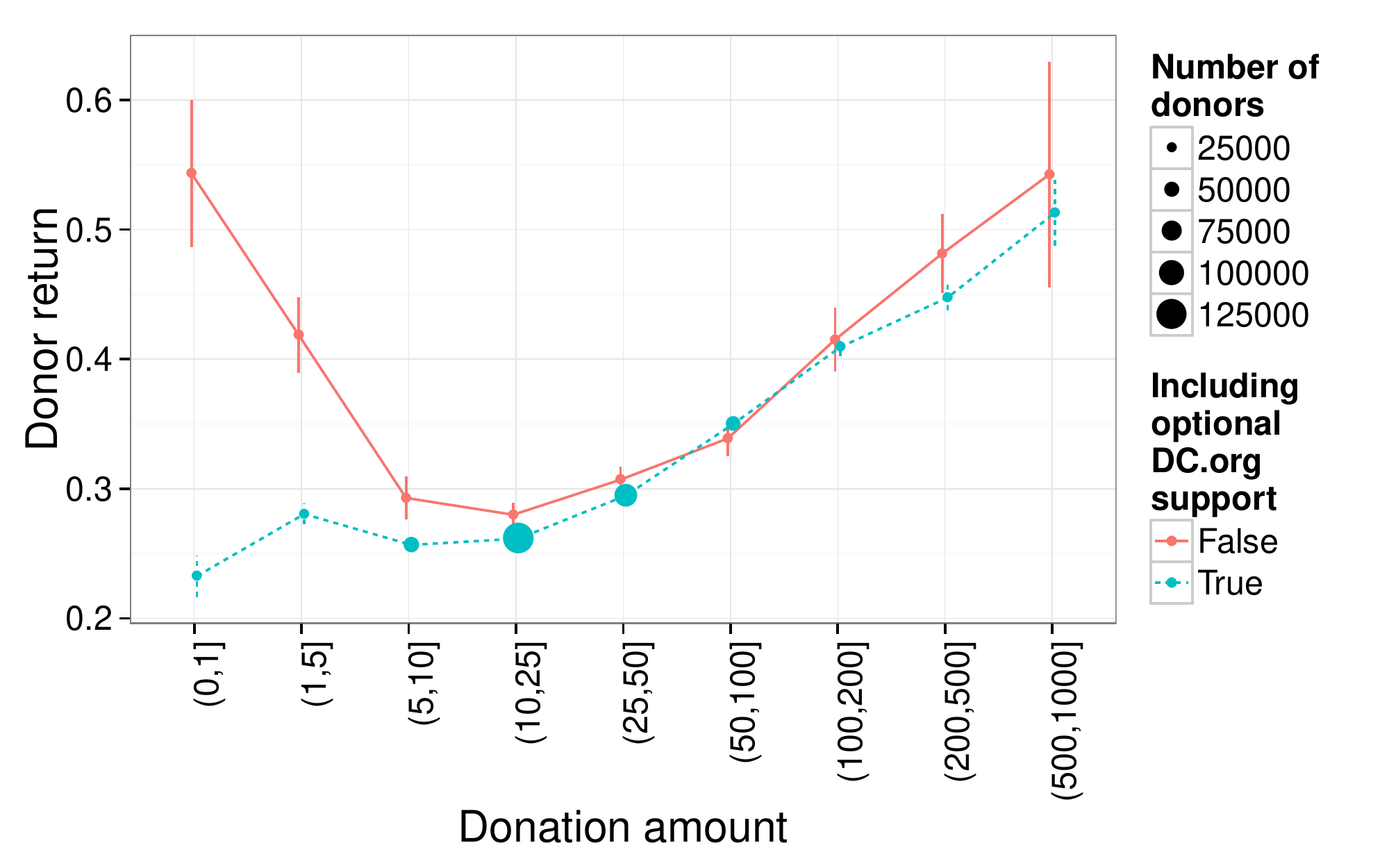}
  \caption{
      Donors expressing commitment to \DC through generous donation amounts are more likely to return.
      Donors opting out of supporting \DC and making small donations are surprisingly loyal to the site (see the text for details).
    }
  \label{fig:return_rate_by_donation_amount_and_opt_donation}
\end{figure}

We show retention rates across different donation amounts in Figure~\ref{fig:return_rate_by_donation_amount_and_opt_donation}.
The first donation amount is highly indicative of donor return. 
This effect is particularly strong for extreme donation amounts (\$100+) that are less common.
It shows that we can use initial donation amounts to predict whether a donor will return in the future.
Note that these high donations amounts often occur to complete a project (see Section~\ref{subsec:donation_position}).

Interestingly, opting out of supporting \DC is not an indicator of donors with low propensity to return.
In fact, donors that opt out are at least as likely to return as their opt-in counterparts across all donation sizes.
For very small donations (\$1-\$5) that do not include \DC support we further observe very high return rates (though this is a relatively small group of donors). 
This counterintuitive finding --- donors with small donations that do not include support for \DC are actually very loyal --- demands further investigation.
Note that we do exclude all donations by teachers who are known to regularly support each other with very small donations
as well as all donations affected by promotions.
We observe the same behavior across teacher-referred and site donors as well.

\begin{table}[t]
\begin{center}
\begin{tabular}{p{14mm}p{9mm}rp{12mm}p{12mm}}
\toprule
\textbf{Donor TR} & \textbf{Donor DPI} & \textbf{\#Donors} & \textbf{Donation Amount} & \textbf{Return Rate}\\
\midrule
Site donor & no PI & 154965 & \$43.70 & 16.4\%\\
Site donor & DPI & 133543 & \textbf{\$67.01} & \textbf{41.1\%}\\
\sectionrule
TR donor & no PI & 87248 & \$40.80 & 10.4\%\\
TR donor & DPI & 95033 & \textbf{\$53.24} & \textbf{31.8\%}\\
\bottomrule
\end{tabular}
\caption{
Return rate and average donation amount across donors that are (not) teacher-referred (TR)
and do (not) disclose personal information (DPI; location or photo).
Disclosure of personal information is strongly correlated with higher donation amounts and return rates
(gains of 20\% and more).
\label{table:return_rate_by_disclosing_information}
}
\end{center}
\end{table}

%

\subsection{Disclosure of Personal Information}
\label{subsec:disclosing_personal_information}
Last, we explore how donor retention is correlated with increased disclosure of personal information.
Sharing more personal information arguably expresses a certain level of trust towards the organization~\cite{HarrisonMcKnight2012trust}.

We define ``discloses personal information'' (DPI) as disclosing location or uploading picture (or both) and measure retention rates across the group of donors that does disclose information and the group that does not.
The results are listed in Table~\ref{table:return_rate_by_disclosing_information}.
For both teacher-referred (TR) and site donors, those that disclose additional personal information (DPI) are much more likely to return.
This effect is very large with differences of over 20\% in both cases (which more than doubles the return rate).
Furthermore, donors that disclose personal information make much larger donations on average.


Overall, these results show that disclosing personal information is correlated with higher levels of loyalty which allows us to exploit this correlation to understand and predict which groups are more likely to return.

\pagebreak 
\section{Teacher Perspective}
\label{sec:teacher}
This section explores donor retention factors around the teachers involvement such as the repeated project efforts by the teacher, teacher-donor communication, and the teacher's use of Facebook for soliciting donations.

\begin{figure}[t]
  \centering
  \includegraphics[width=1.0\linewidth]{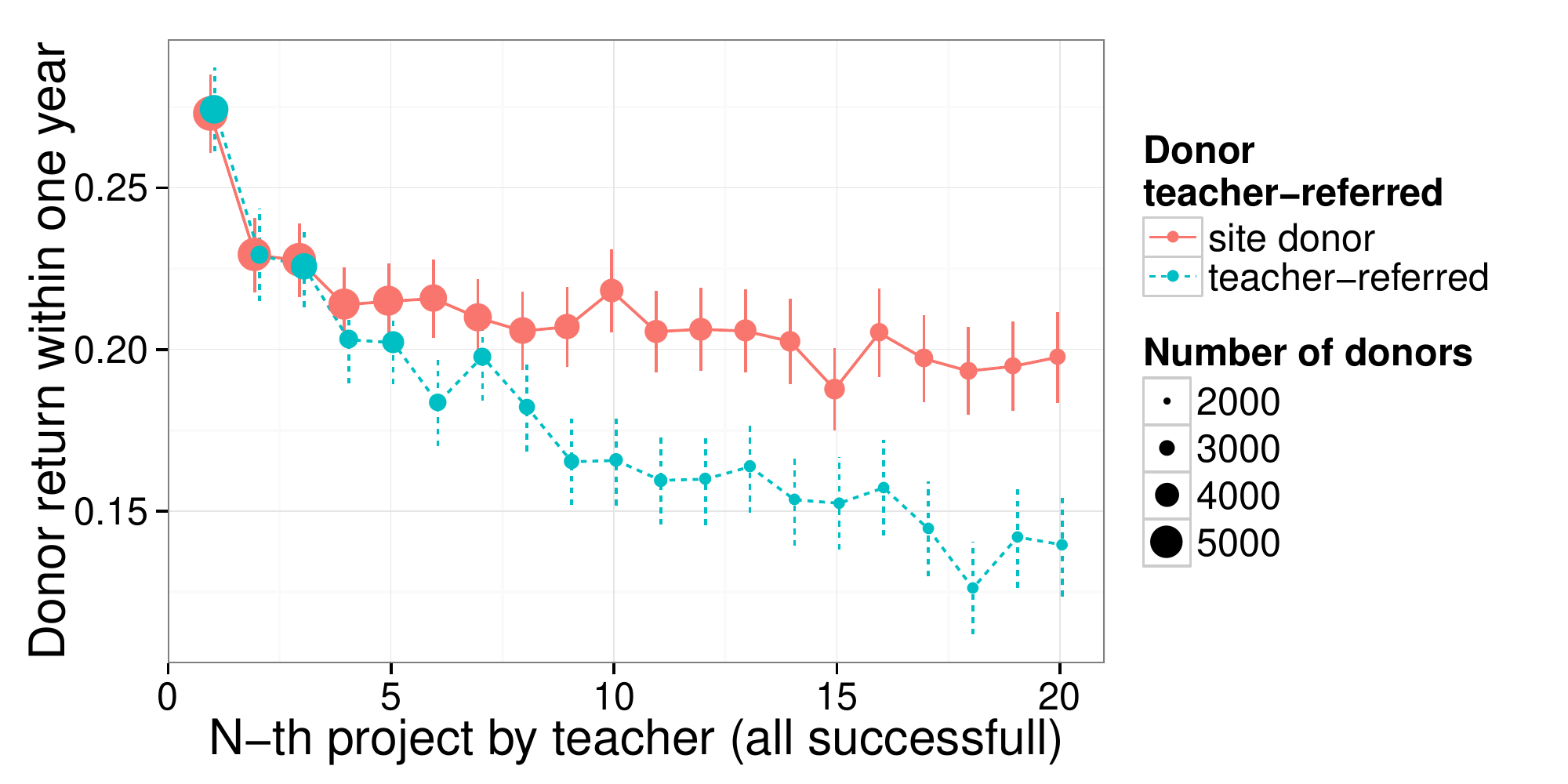}\\
    \includegraphics[width=1.0\linewidth]{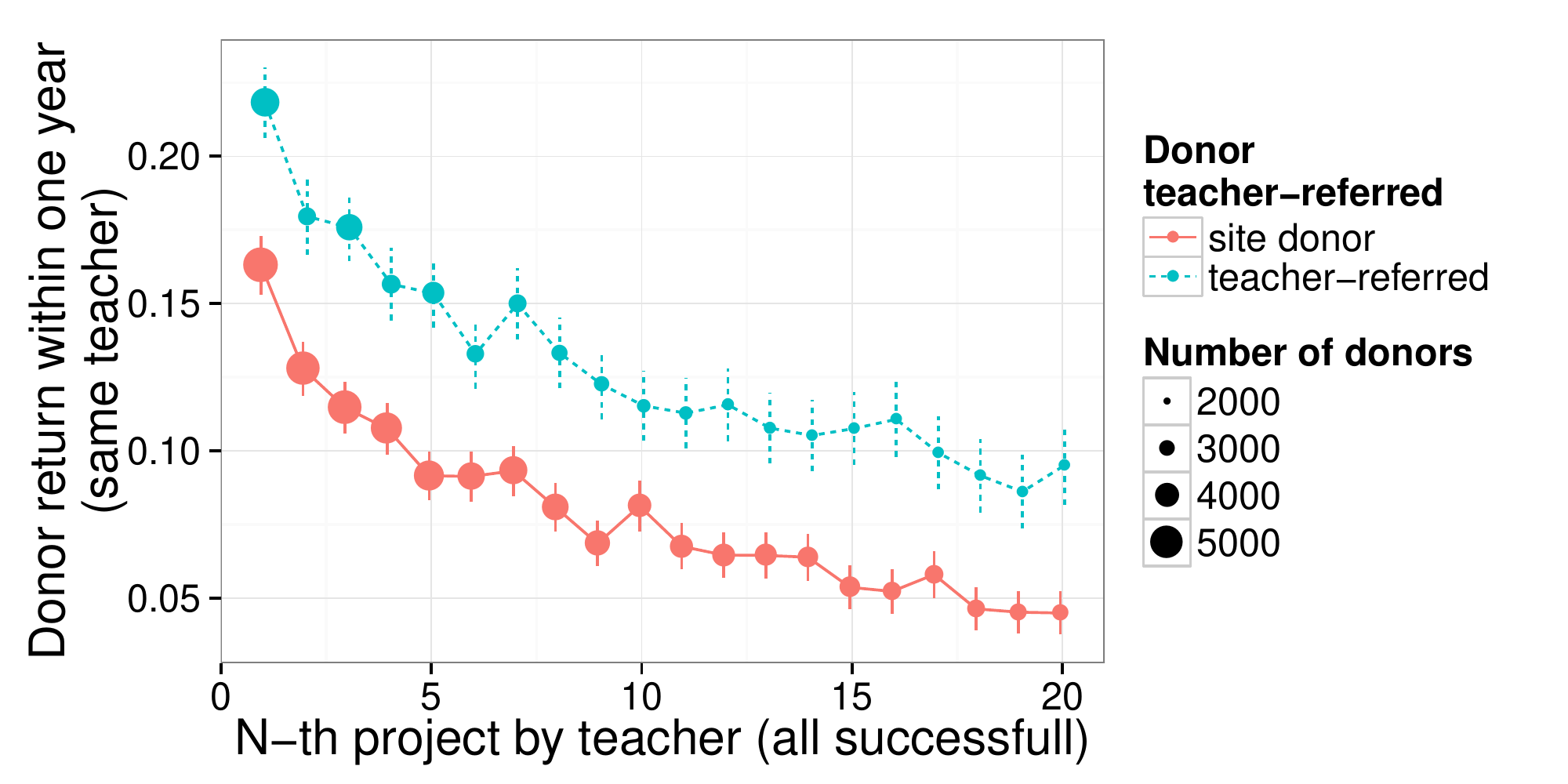}
  \caption{
      (Top) Teacher-referred donors are less likely to return to the site as the teacher posts new projects over time.
      For site donors, the effect levels off after an initial decrease in retention rate.
      (Bottom) Both teacher-referred and site donors are less and less likely to return to the same teacher over the ``teacher lifetime''.
    }
  \label{fig:return_rate_by_nth_project_teacher}
\end{figure}

\subsection{Expertise -- Do Teachers Become More Successful Over Time?}

We seek to empirically answer the question whether experienced teachers become more successful in retaining donors---both for the site and for their own projects---by measuring return rates for the teacher's first project, second project, and so on.
To avoid bias of failed projects we restrict ourselves to only successful projects for this analysis and only consider teachers that posted at least 20 projects.
As donors to earlier projects have more time to return compared to donors to later projects, we also require donors to return within one year.

The results are shown in Figure~\ref{fig:return_rate_by_nth_project_teacher} (top: return to site; bottom: return to teacher).
We observe the opposite from what one might expect.
Experienced teachers do not become more effective at retaining donors.
Instead, they are most successful in retaining donors (to \DC; top plot) in their first projects.
After those, the return rates are monotonically decreasing over time.
However, we see that the effect levels off for site donors whereas the retention continues to decrease for teacher-referred donors.
This could be explained by viewing teacher-referred donors as a limited resource available to the teacher. 
Teachers are only able to receive a certain amount of donors from their personal support network.
This support is limited and asking over and over again for new projects is less and less likely to be successful as the teacher ``drained'' most of the resources available to them already.
On the other hand, site donors are a much larger group which could explain why the red curve for site donors is leveling off instead of decreasing.

The bottom plot in Figure~\ref{fig:return_rate_by_nth_project_teacher} shows return rates to the same teacher.
Here, we observe decreasing retention rates for both teacher-referred and site donors.
Note, however, that teacher-referred donors are now more likely to return, which is to be expected as they were referred by the teacher themselves.
This picture seems somewhat discouraging for teachers looking for continued support for their classrooms.
As donor retention becomes more and more challenging for the teacher,
they are likely to be less successful over time (and in fact, that is what we observe in the data).

What are possible explanations for this negative trend?
\DC wants teachers to be successful, in particularly new teachers, and has implemented several features to support this.
The main search interface for projects contains a filter option ``never before funded teachers'' (introduced in 2008, so before the start of our observation period).
Furthermore, each project page lists the teacher's number of previous successful projects, making it easy for donors which are passionate about funding new teachers to direct their support.
This suggests that donors are somewhat ``fair'' in distributing their support
as they seem to favor supporting a new teacher instead of funding the 20th project by another.




\subsection{Giving Thanks -- Appropriate Recognition of Donors}
\label{subsec:confirmation_note}

\begin{figure}[t]
  \centering
  \includegraphics[width=1.0\linewidth]{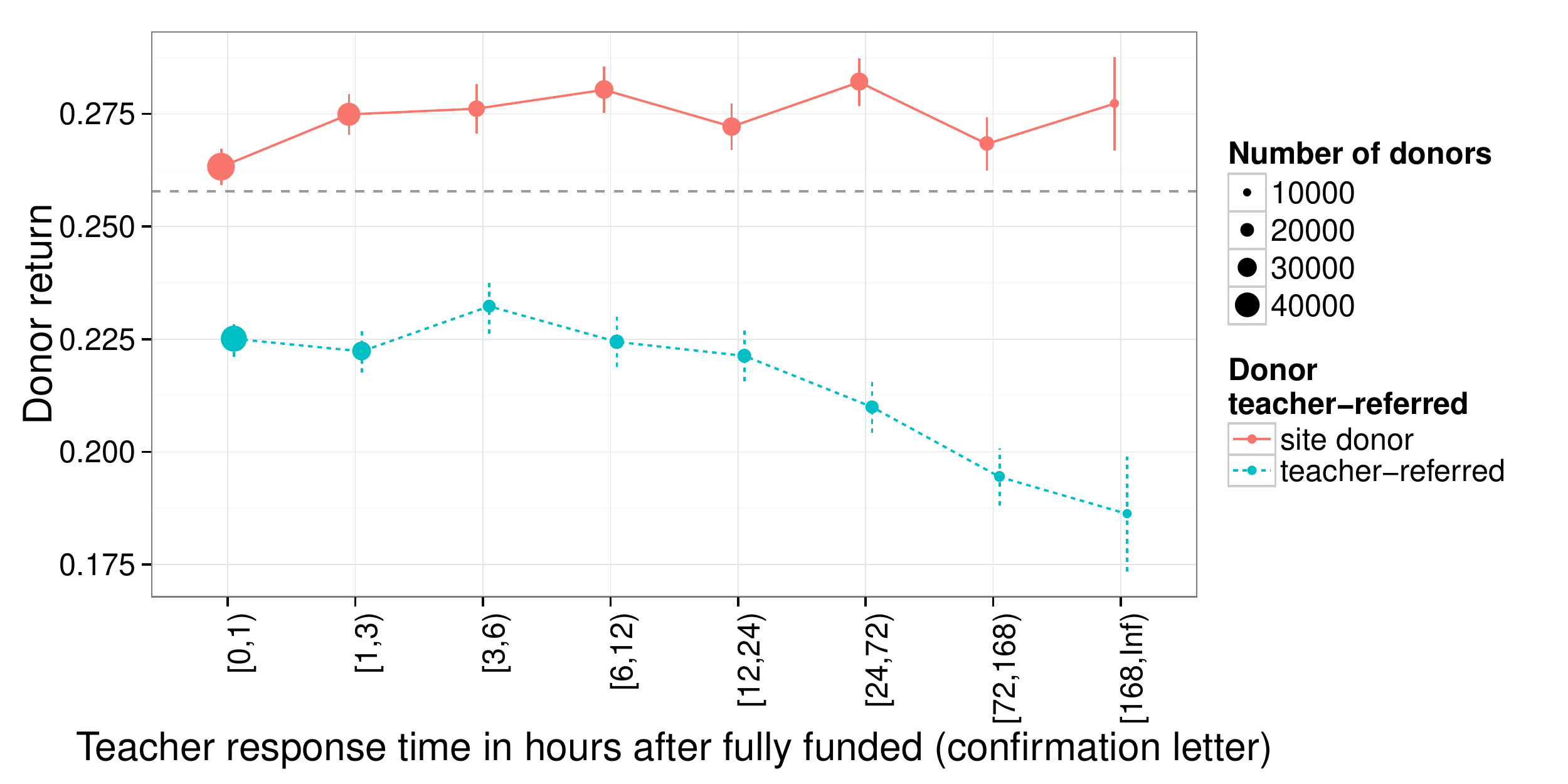}
  \includegraphics[width=1.0\linewidth]{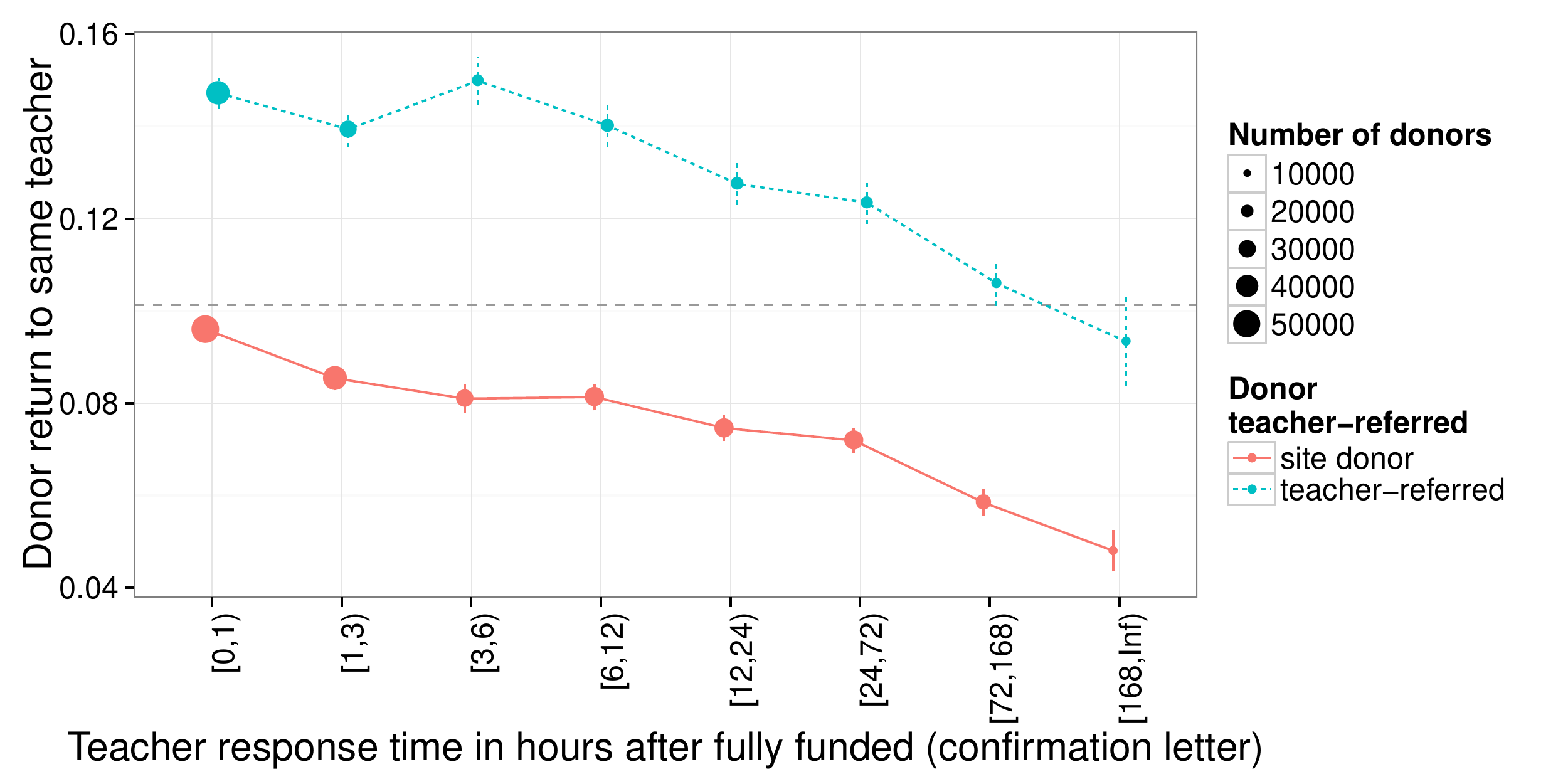}
  \caption{
    Effect of giving thanks through a confirmation note right after project becomes fully funded.
    Top: Donor return to site. 
    Bottom: Donor return to same teacher.
    The dashed lines represent the respective average return rates.
    }
  \label{fig:return_rate_by_teacher_confirmation_letter_response_time}
\end{figure}

Next, we quantify the effect of timely thank you messages on donor retention.
As described in Section~\ref{sec:dataset}, teachers send out a ``confirmation note'' after the project becomes fully funded thanking their donors.
We find that practically every teacher writes such a note eventually.
Fundraising literature as well as anecdotal evidence from practitioners highlight the importance of providing appropriate recognition to the donor~\cite{sargeant2008donor}.
Failure to do so might lead to a lowering of future support or its complete termination~\cite{boulding1973economy}.

%
We measure retention rates to the site and the same teacher across confirmation note response time in hours as shown in Figure~\ref{fig:return_rate_by_teacher_confirmation_letter_response_time}.
For site return (top) we only observe an effect for teacher-referred donors where slower response times show lower retention rates.
In particular, response times within the first 24 hours are correlated with significantly higher return rates.

To sum up, donor retention on teacher level shows larger effects of response time for both teacher-referred and site donors. The effect is more pronounced for teacher-referred donors which suggests that they have higher expectations to be thanked or are more sensitive to hearing back promptly.


\subsection{Communicating Impact -- Let Donors Know The Difference}
Similarly to thanking donors for their support (see Section~\ref{subsec:confirmation_note}), communicating impact has been identified as an important driver of donor loyalty~\cite{sargeant2008donor}.
As introduced in Section~\ref{sec:dataset}, \DC asks teachers to write an ``impact letter'' to their donors for exactly these reasons after they have received the donated material.
These impact letters usually include photos of students using the recently donated materials.

\begin{figure}[t]
  \centering
  \includegraphics[width=1.0\linewidth]{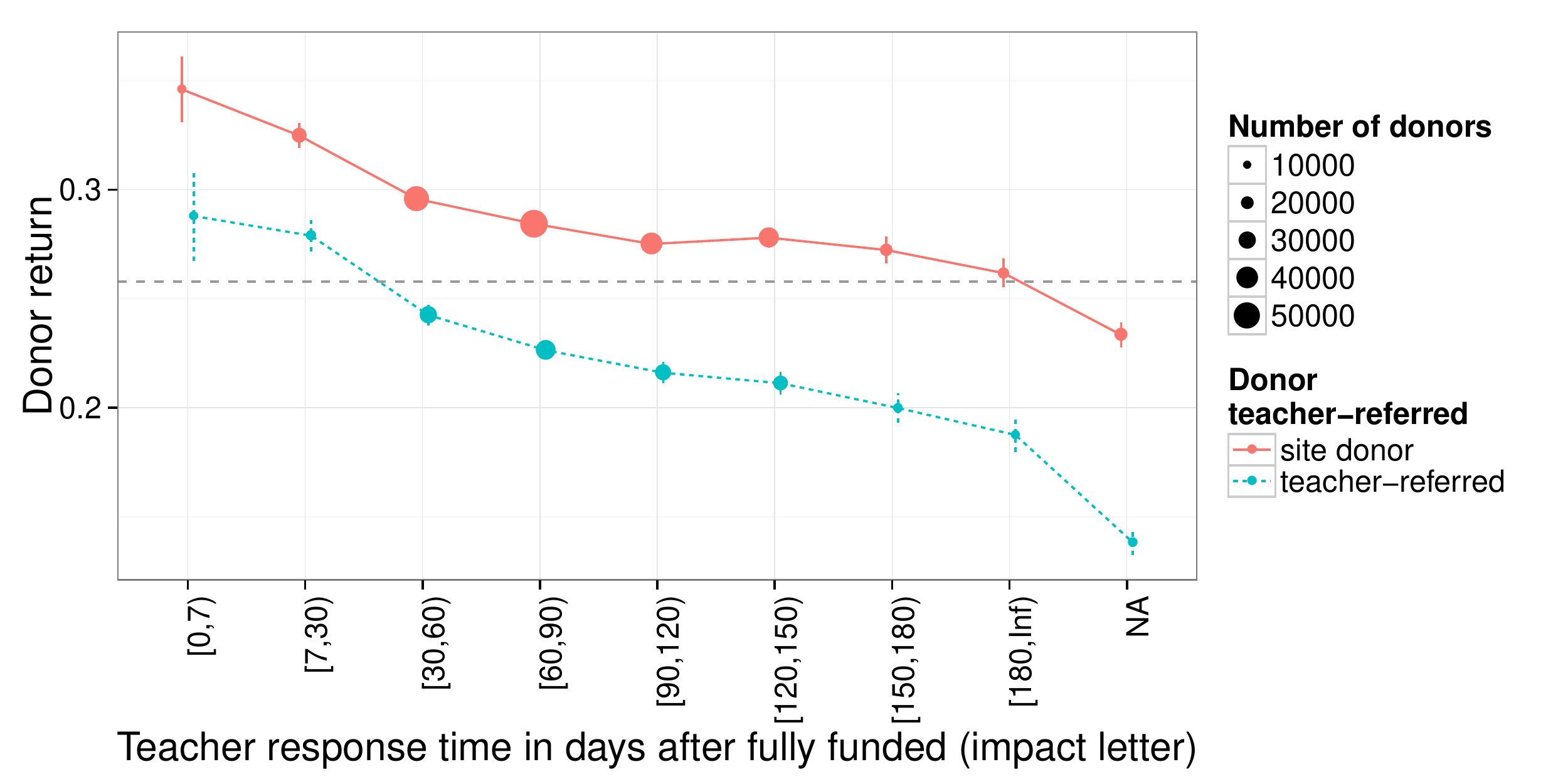}
  \includegraphics[width=1.0\linewidth]{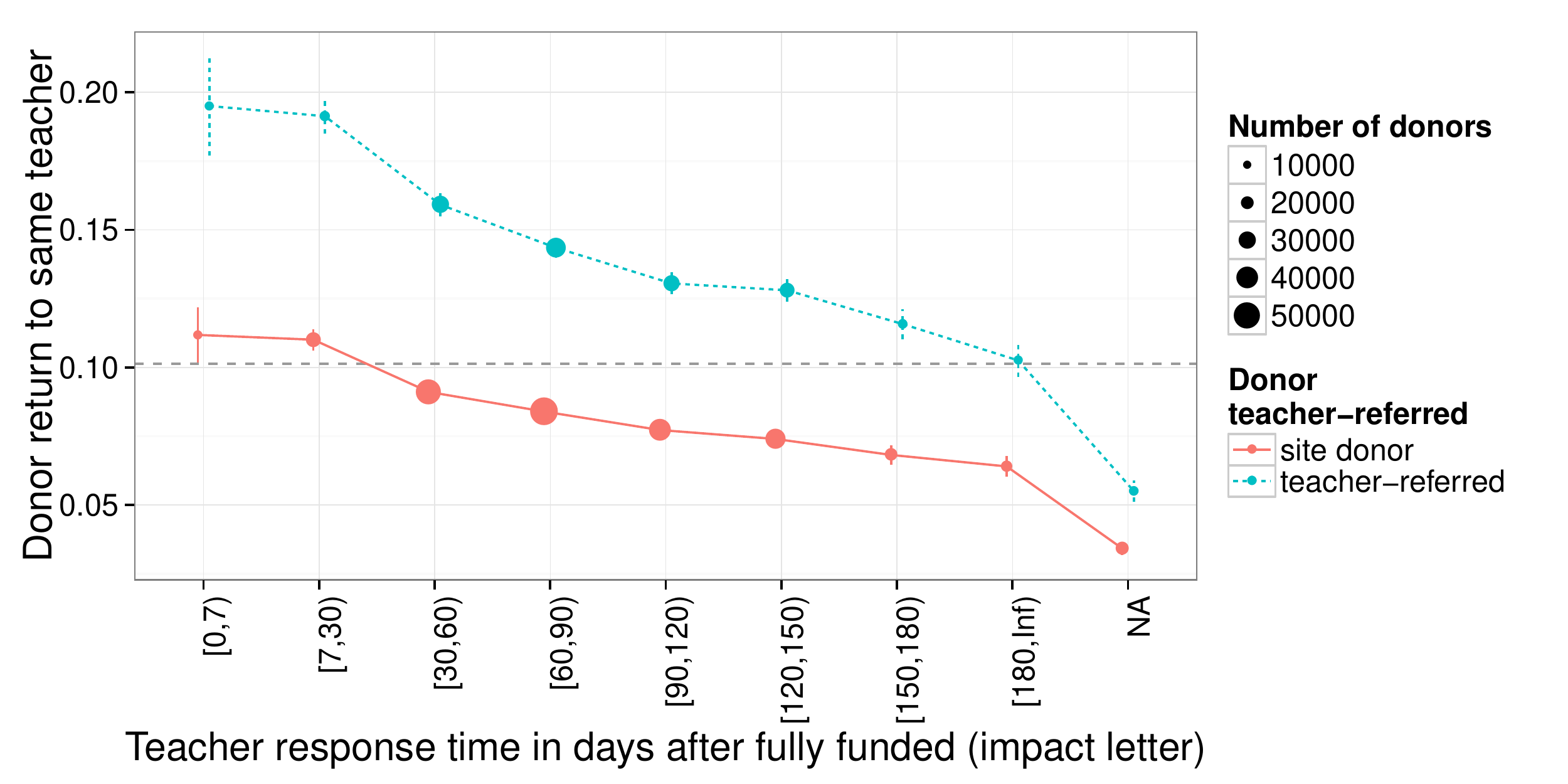}
  \caption{
    Effect of communicating impact through an impact letter.
    Top: Donor return to site. 
    Bottom: Donor return to same teacher.
    In all cases, timely communication is very strongly correlated with donor return.
    The dashed lines represent the respective average return rates.
    %
    }
  \label{fig:return_rate_by_teacher_impact_letter_response_time}
\end{figure}

Similar to the previous section, we analyze the effect of timely response rates (in days) as well as failing to submit an impact letter on the teacher's part.
The results are shown in Figure~\ref{fig:return_rate_by_teacher_impact_letter_response_time}.
In short, communicating impact to donors is very important for both retention to site (top) and to the same teacher (bottom) and for both teacher-referred donors as well as site donors.
We observe strictly decreasing retention rates for longer response times with failure to submit an impact letter (NA) being the lowest.
All differences in return rates are very large with return to the same teacher for teacher-referred donors again being the most pronounced.

\subsection{Growing Your Support Network Through Social Media}

\begin{figure}[t]
  \centering
  \includegraphics[width=1.0\linewidth]{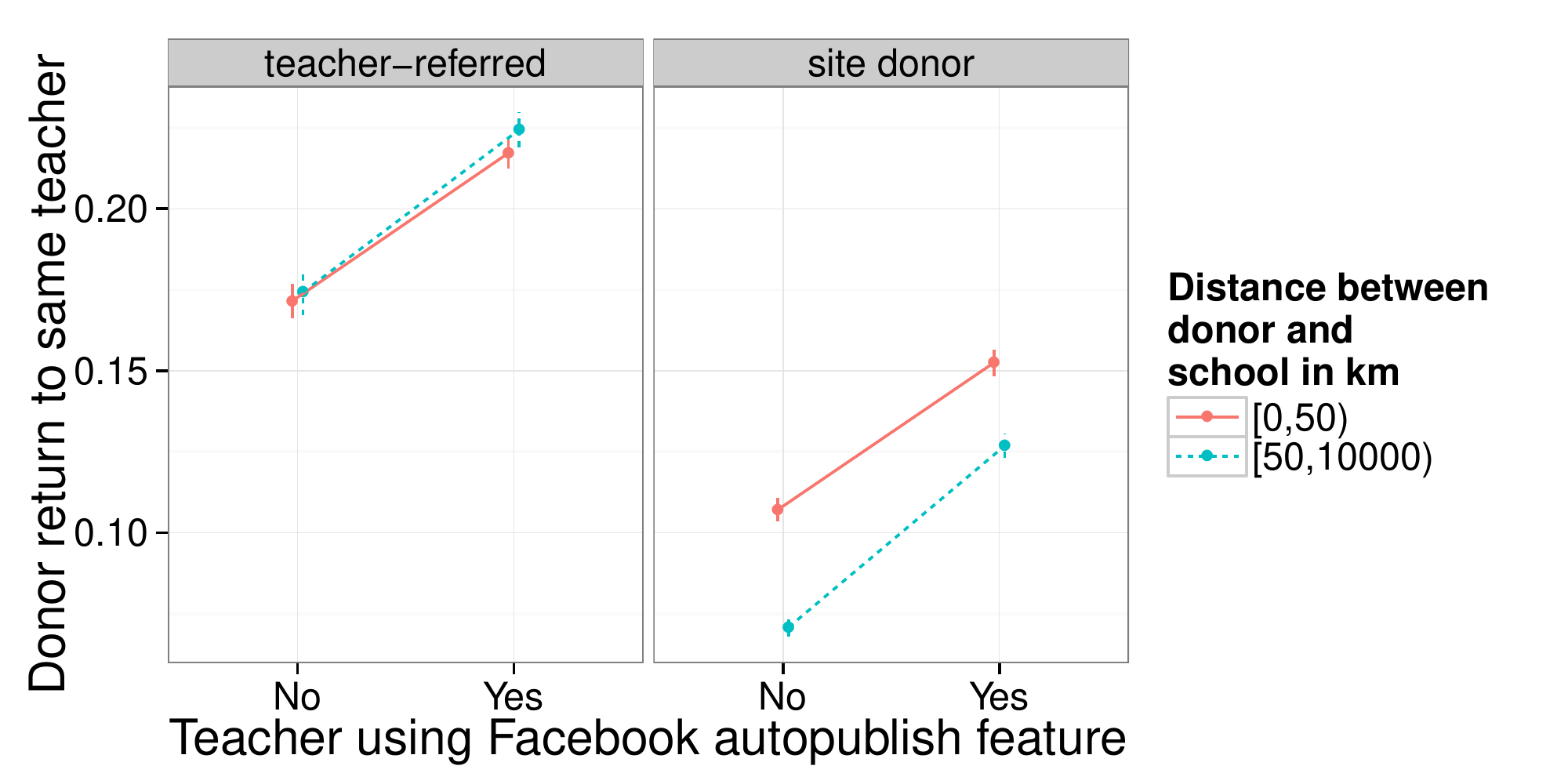}
  \caption{
    Donors that donate to teachers who allow \DC to publish posts on their behalf are significantly more likely to return.
    However, the effect persists for both local and distant as well as teacher-referred and site donors suggesting that this is not a causal effect (see text for details).
    }
  \label{fig:return_rate_by_teacher_facebook}
\end{figure}


\DC offers teachers to automatically post key events about their project (first posted, first donation, \etc) on Facebook on their behalf. 
While we cannot know for sure the degree to which teachers make use of social media to raise support for their projects, surely the ones that opt in to use this autopublishing feature will regularly present friends and followers with updates on the teacher's fundraising efforts.
Do teachers that opt in have higher return rates? 


We show return rates in Figure~\ref{fig:return_rate_by_teacher_facebook}. Donors that donate to ``opt-in teachers'' are more likely to return. However, this effect is identical for local and distant donors, and even more importantly exists for both teacher-referred and site donors 
(though we would expect that site donors are not connected to the teacher on Facebook). 
This strongly suggests that the observed effect is not causal (\ie, due to the use of the Facebook feature) but that these might be a more fundamentally different group of teachers that is better at retaining donors for different reasons. Perhaps, teachers that opt in are generally more involved in soliciting donations from their peers.

This analysis cannot replace experiments specifically designed to measure the effect of social media use on support network within the crowdfunding community. However, since opting in is yet another signal that distinguishes returning donors from those that do not return it can still be helpful in predicting donor return.



\section{Summary of Results}
\label{sec:summary}

In this Section we briefly summarize our findings on donor retention factors around the 
project (Section~\ref{sec:project}),
donor (Section~\ref{sec:donor}),
and teacher (Section~\ref{sec:teacher}).
\vspace{-1.5mm}
\begin{itemize}[noitemsep]
  \item \textbf{Project}: 
    Donors experiencing a successful first project, small projects in particular, are more likely to return.

  \item \textbf{Donor}: 
    Teacher-referred donors tend to be local and start off projects with early donations.
    Site donors fund projects much farther away and are more likely to finish off projects in which case they are very likely to return.
    Donors who give extraordinary amounts or disclose optional personal information about themselves are particularly loyal to \DC.

  \item \textbf{Teacher}: 
    Teachers are less successful over time in retaining donors.
    Timely recognition of donations as well as communicating the donor's impact is crucial, particularly for return to the same teacher and teacher-referred donors.
\end{itemize}

\section{Predicting Donor Return}
\label{sec:prediction}
\begin{table}[t]
\centering
\begin{tabular*}{\columnwidth}{lp{6cm}}\toprule
  \textbf{Feature Set} & \textbf{Features} \\
  \midrule
  Time (1) & month of donation \\
  Project (2) & eventual project success, project cost \\
  Donor (10) & donation amount, donation includes optional support, 
distance to school, donation position (first, middle, last), donor photo published,
donor teacher-referred, donor asked for student thank you notes \\
  Teacher (8) & $n$-th project by teacher, completion and response time for confirmation note, impact letter, and student thank you notes,
use of Facebook autopublish feature \\
\bottomrule
\end{tabular*}
\caption{
We consider four different categories of features as well as their combinations. 
The number in parenthesis denotes the number of features in the group.
}
\label{tab:feature_list}
\end{table}

Next, we build on insights from previous sections to predict donor retention on an individual level using standard machine learning techniques.

\xhdr{Features used for learning}
We define a series of models that use different sets of features based on the factors explored in previous sections. We focus on four types of features:
\vspace{-1.5mm}
\begin{itemize}[noitemsep]
	\item {\bf Time:} We simply include the time of donation to control for temporal effects, like the state of the U.S. economy and changes in donor population.
	\item {\bf Project:} Based on our analysis in Section~\ref{sec:project} we describe the project with its cost and whether it succeeded eventually.
	\item {\bf Donor:} In Section~\ref{sec:donor} we found features of the donor, like the geographic distance from the school and the donation position within the project to be important.
	\item {\bf Teacher:} Insights gained in Section~\ref{sec:teacher} demonstrate that properties of the teacher and communication with the donor to be important as well.
\end{itemize}
Table~\ref{tab:feature_list} gives a complete list of features. We include binned variants of donation amount and project cost and standardize all features to have zero mean and unit variance.

\xhdr{Experimental setup}
We report performance of Logistic Regression models though Random Forest and SVM models gave very similar results.
Because of the unbalanced dataset (74.6\% of donors did not return for a second donation) and the trade-off between true and false positive rate associated with prediction we choose to compare models using the area under the receiver operating characteristic (ROC) curve (AUC) which is equal to the probability that a classifier will rank a randomly chosen positive instance higher than a randomly chosen negative one.
Thus, a random baseline will score 50\% on ROC AUC.
We estimate ROC AUC through 10-fold cross-validation across the full dataset of 470,789 first-time donors.
We experimented with weighting samples inversely proportional to class frequencies in the training set to address the class imbalance in the dataset and observed slight boosts in predictive accuracy at the expense of worse model calibration.

\begin{table}[t]
\begin{center}
\begin{tabular}{lr}
\toprule
\textbf{Model}  & \textbf{ROC AUC}  \\
\midrule
Random Baseline & 0.50\\
\sectionrule  
Time (t)        & 0.53\\
Project (P)     & 0.54\\
Donor (D)       & 0.72\\
Teacher (T)     & 0.55\\
\sectionrule  
t + P           & 0.56\\
t + P + D       & 0.73\\
t + P + D + T   & \textbf{0.74}\\
\bottomrule
\end{tabular}
\caption{
Performance results for predicting donor return after observing the first donation only.
Reported numbers are based on a Logistic Regression model on the full dataset (25.4\% donor return probability) through 10-fold crossvalidation.
}
\label{table:prediction_results}
\end{center}
\end{table}

\xhdr{Summary of results}
The results are given in Table~\ref{table:prediction_results}.
As reported in Section~\ref{subsec:donor_retention_on_dc}, donor retention has been decreasing over time.
The Time model exploits this correlation and performs at 0.53 ROC AUC.
Project features (success and cost) perform slightly better at 0.54 ROC AUC.
%
The model based on donor features already performs quite well at 0.72 ROC AUC (note it is also the largest group of features).
%
Teacher features perform slightly better than the Time or Project model at 0.55 ROC AUC.

Overall, we learn that the donor features are the most predictive of the return outcome.
With about 0.72 AUC they help in distinguishing between returning and non-returning donors correctly.
We further explore different feature set combinations and see performance improvements of 
0.56 for Time + Project, 0.73 when adding Donor features, 
and finally up to 0.74 for the ``full'' model combining all features.
This means that given two first-time donors, one returning and one non-returning, the model is able to pick the donor that is more likely to return in about 74\% of all cases.

\xhdr{Exploring the model structure}
Inspecting the full model for the largest absolute feature weights (though we caution about any absolute interpretation of importance) reveals that the model strongly relies on the following features:
The model uses extraordinary donation amounts to infer higher donor return rates.
As expected, smaller projects further increase the predicted donor return rate (see Section~\ref{subsec:project_cost}).
In particular, the model relies more strongly on distance; that is, whether the donor is local (under 25km) or distant (1000km or more), or whether donor location was not disclosed.
The predicted return probability is also strongly influenced by whether the donor was teacher-referred, whether the donor uploaded a photo, and whether the donor made the completing donation to the project.
The model further puts particular emphasis on whether the teacher failed to upload thank you photos and how many projects the teacher has had before the current one.
Lastly, the model exploits the temporal trend to downweight recent donors.


\begin{figure}[t]
  \centering
  \includegraphics[width=1.0\linewidth]{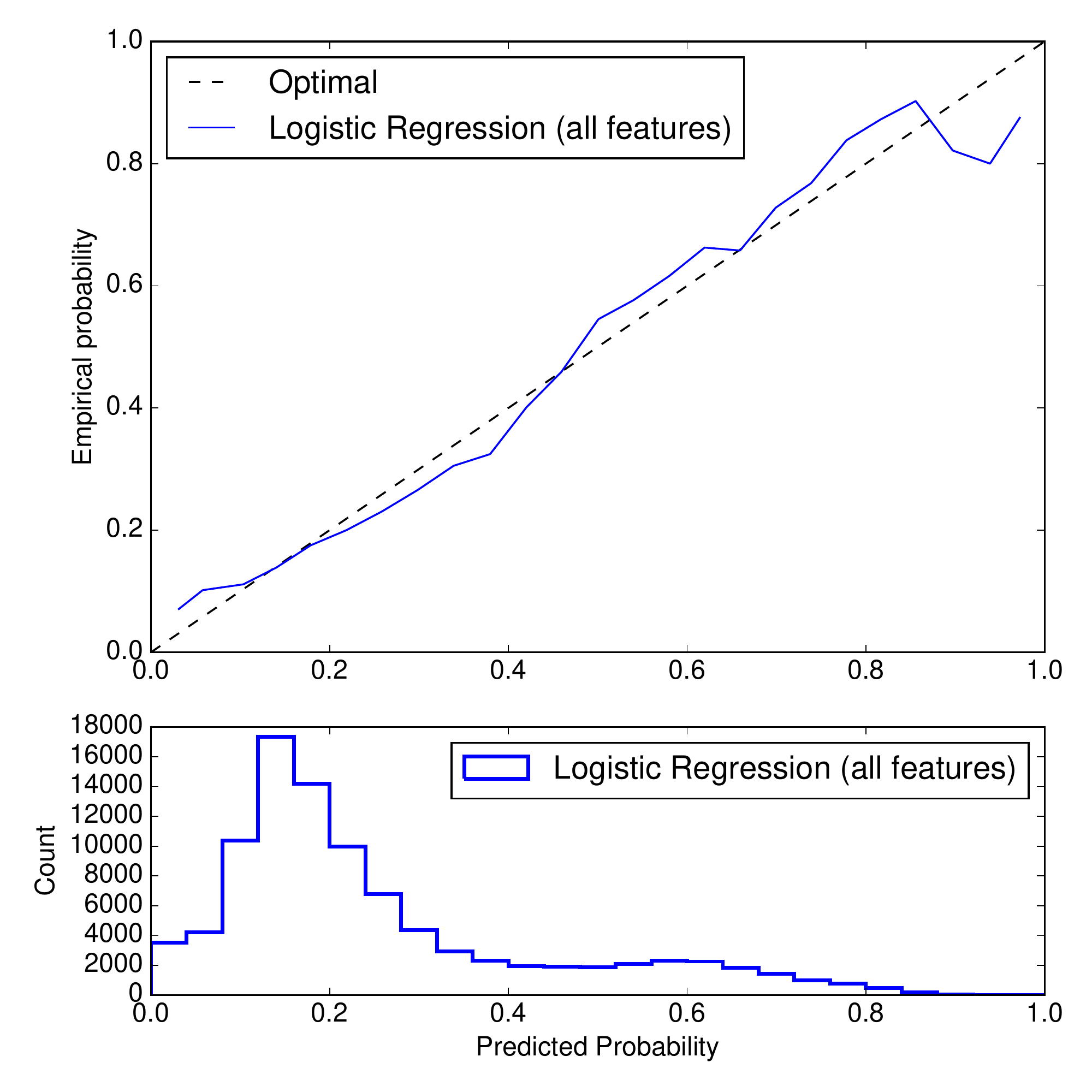}
  \caption{
    Model calibration plot of predicted probabilities of donor return against empirical probabilities showing that the Logistic Regression model using all features predicts well-calibrated probabilities.
    }
  \label{fig:calibration_plot}
\end{figure}

\xhdr{Model calibration}
We plot return probabilities as predicted by our full model against empirical probabilities found in the test data in Figure~\ref{fig:calibration_plot}.
The Logistic Regression model using all features is well-calibrated 
(\ie, predicts sensible donor return probabilities) and only diverges at the extremes for which we only have very few examples.
Interestingly, the histogram of predicted probabilities reveals that there is a large population that is unlikely to come back (left) as well as a smaller population with high probability of return (right).
It might be much harder and more costly to convince unlikely-to-return donors to make another donation compared to donors that are likely to return.
To be more cost-effective, we therefore propose to concentrate marketing efforts on the latter group.

\xhdr{Discussion}
Almost certainly, there is much room for improvement in predicting donor return on an individual level for example through additional features (\eg, donor log-in history, content analysis of project essays, donor, and thank you messages) or through more powerful models.
We see these results as an encouraging proof-of-concept that we can learn so much about donors, using information limited to only their first donation, that we can predict with reasonably high accuracy which donor is going to return for a second donation.
Such models could prove to be very useful in informing fundraising campaigns and providing a basis for modeling effects on donor retention through targeted interventions.
Perhaps, campaigns should focus more on the subpopulation that is more likely to return (cf. Figure~\ref{fig:calibration_plot}).



\section{Related Work}
\label{sec:related}

We discuss related work in two parts. 
First, we mention the work that focuses on online crowdfunding platforms, which mostly deals with predicting success of funding campaigns. 
Second, we discuss survey-based research on donor retention in offline charities and non-profits.\

\xhdr{Online crowdfunding platforms}
Emergence of crowdfunding platforms, like Kickstarter, Kiva, and Prosper, has lead to a rich line of work on quantifying dynamics of funding campaigns~\cite{ceyhan11prosper,Collier2010Prosper,flannery2007kiva,Greenberg13Crowdfunding,Hui14Community,liu2012loan}. For example, studies have examined intrinsic dynamics of projects~\cite{Mollick14Dynamics}, how entrepreneurs use crowdfunding platforms~\cite{Hui14Community}, and how these platforms compare among themselves~\cite{Greenberg13Crowdfunding}.
Research has also sought to develop tools that would help crowdfunding project creators~\cite{Greenberg13Tools} by predicting the probability of the project being successfully funded~\cite{Greenberg13Tools,kuppuswamy2013crowdfunding,Rao14Dynamics} and by predicting the number of contributions the project will eventually receive~\cite{Lu14Media}. 
Other works have investigated the impact of factors like distance, promotional activities, project updates, and the use of social media on the success of projects~\cite{agrawal2011geography,Lu14Media,Muller14Distance,Xu14Money}. Common to all these works is that they investigate the dynamics of crowdfunding platforms from the viewpoint of the project and the project creator. The central question around these works is how to identify best practices and help project creators to get their projects eventually funded.

In contrast, our work does not focus on the dynamics of projects. Rather, we investigate the dynamics of donors. We examine the dynamics of crowdfunding platforms from the viewpoint of the donor and quantify donor behaviors that are indicative of donor's return and continued involvement with the crowdfunding platform. Another difference is that majority of works have investigated crowdfunding in the context of raising money for commercial projects. 
There, investors expect some return either by charging interest rate for the money they lend (as in the case of microlending platform Prosper) or by pre-ordering the product (as in the case of Kickstarter). 
In contrast, our work here examines charitable contributions where investors (\ie, donors) do not expect any tangible return beyond the successful completion of the project (except perhaps the acknowledgement of their support or tax deductions). 

\xhdr{Donor retention in charities}
A rich line of research on traditional offline charities has emphasized high rates of donor attrition as well as the importance of donor retention for charities to achieve their goals~\cite{barber2013donor,sargeant2008donor}. 
Using survey-based methodology, researchers have studied various factors related to donor retention~\cite{bennett2006predicting}. For example, importance of acknowledging the donor by saying ``thank you'' has been recognized as an important factor~\cite{merchant2010don,veldhuizen2010thank}. Furthermore, content analysis of arguments used in fundraising letters revealed that fundraisers tend to use emotional arguments more than logical ones~\cite{ritzenhein1998content}. Other important factors for donor retention include relationship building, communicating impact, trust, commitment, satisfaction, and involvement~\cite{naskrent2011influence,sargeant2001relationship,sargeant2008donor,sargeant2006perceptual}.

Perhaps the most related to our work here is the research that investigates the roles that computational technology plays in support of non-profit fundraising~\cite{goecks2008charitable}. In particular, recent work used the \DC dataset in order to examine the value of completing crowdfunding projects and found that completing a project leads to larger donations and increased likelihood of returning to donate again~\cite{wash2013value}. Our work builds on this line of work and examines complete \DC data in order to better understand donor attrition and identify means for increasing donor retention.

Our work further relates to the broader area of contributor retention in various online settings including newsgroups \cite{arguello2006talk,joyce2006predicting}, forums \cite{lampe2005follow}, Q\&A sites \cite{dror2012churn,pudipeddi2014user,yang2010activity}, Wikipedia \cite{halfaker2012rise}, and social networks \cite{karnstedt2011effect}.
We envision that our study could generalize to these settings as well by contributing proxies for the user's initial motivation and commitment as well as the dynamics around the recognition of their support.



\section{Discussion \& Conclusion}
\label{sec:conclusion}

\xhdr{Summary}
Online crowdfunding platforms face the same challenge of maintaining a relationship with their donors as traditional non-profit organizations do. The present paper takes a first step towards addressing this challenge by analyzing donor behavior within a large crowdfunding platform \DClong. In particular, we focus on first-time donors, the group for which the attrition is by far the largest. We identify a set of factors related to donor retention from project, donor, and teacher (project starter) perspectives and quantify the effects of these factors on donor retention, both to the site and to the individual teacher.

Using just the first donation interaction we show that we can learn a lot about the donor behavior to successfully predict the donor's propensity to return and make further donations. 
In particular, we learn about the donor's initial commitment through the means with which she enters the site (teacher-referred or not), their proximity to the project they are supporting, the amount they are giving, and whether they disclose personal information. 
We have proxies for the donor's sense of impact and trust in the organization through the project cost and size, whether the project is successful, and whether the donor receives a personal letter communicating the impact they have had on the project. Lastly, factors such as timely writing ``thank you'' notes to the donors, teacher experience on the site, or use of social media for solicitation are also correlated with the teacher's ability to retain donors.

Ideally, these factors would represent causal effects to help us understand how to improve donor retention. However, this is not a necessary requirement for such factors to be useful in machine learning models that predict whether or not donor is likely to return. We show that even simple models can predict donor return with reasonably high accuracy. Such models could prove to be very useful for crowdfunding platforms as well as non-profit organizations to efficiently target fundraising campaign efforts.

\xhdr{Implications}
Our findings also inform steps to improve donor retention in crowdfunding communities and non-profit organizations.
For example, our research suggests what factors are important in devising interventions and campaigns targeted at specific donor subpopulations.
We showed that site donors and teacher-referred donors are very different in their behavior and campaign efforts and recommender systems should treat these two groups differently and expect different outcomes. Similarly, some donors prefer to support their local neighborhood whereas other donors are happy to help just about anywhere. This is valuable information for charitable organizations that should direct the flow of donations in a way that maximizes success for their organization and their users. Surely, this will involve many trade-offs (\eg, focusing on first-time vs. high-profile donors or teachers) and this requires future work informing such decisions.

Furthermore, we hope that this work could serve crowdfunding communities and non-profit organizations as an encouragement to start collecting similarly valuable information about their donors and interactions in order to increase their fundraising efficiency.
Even small increases in donor retention can have significant financial impact as donors keep donating for longer periods, very often increase their donation amount, help recruit new donors, and because of the potential savings in marketing to acquire new donors.
We estimate that increasing donor retention by 10\% in the case of \DC would lead to an over 60\% increase\footnote{Increasing donor retention 30\% to 40\% for new donors and 55\% to 65\% for existing donors and using increasing average donation amounts as estimated from the \DC data.}
in obtained donations (or an additional \$15M assuming 100k donors).


\xhdr{Future work}
Future work involves research on how prediction models could inform fundraising strategies (\eg, through field experiments). We believe that a content analysis of project essays, donation messages, and thank you letters could further inform our understanding of the donor-teacher relationship. This paper could also be complemented by qualitative evidence from donors and teachers as well as online field experiments to examine which subgroups of donors to target and how to target them in order to increase donor retention rate. In designing such experiments, one should be mindful of ethical issues as interventions in this space could have negative impact on often already disadvantaged public school classrooms.
Future work should further investigate how this work fits into the broad area of crowdsourcing and what retention factors prove to be valuable across a variety of platforms.

It is our hope that the present work will be able to serve as a basis for more research on maintaining donor relationships in online crowdfunding platforms as well as offline non-profit organizations that aim to improve their ability to retain donors for good causes.

\xhdr{Acknowledgments}
We thank Vlad Dubovskiy and Thomas Vo at DonorsChoose.org for facilitating the research, Rok Sosi\v{c} for many helpful discussions, and the anonymous reviewers for their valuable feedback.
This research has been supported in part by NSF
IIS-1016909,              
CNS-1010921,              
IIS-1149837,       
IIS-1159679,              
ARO MURI,                 
DARPA SMISC,            
Boeing,    
Facebook,
Volkswagen,                 
Yahoo, and
Stanford Data Science Initiative.

\pagebreak
\balance
\bibliographystyle{abbrv}
\bibliography{refs}


\end{document}